\documentclass{sigchi}

\def\authorversion{}
\def\emptyauthor{}

\toappear{\scriptsize 
\ifdefined\authorversion
This is the authors' version of the work. It is posted here for your personal use. Not for redistribution. The definitive Version of Record was published in: \\
\else
Permission to make digital or hard copies of all or part of this work for personal or classroom use is granted without fee provided that copies are not made or distributed for profit or commercial advantage and that copies bear this notice and the full citation on the first page. Copyrights for components of this work owned by others than ACM must be honored. Abstracting with credit is permitted. To copy otherwise, or republish, to post on servers or to redistribute to lists, requires prior specific permission and/or a fee. Request permissions from permissions@acm.org. \\
\fi
{\emph{CHI '20, April 25--30, 2020, Honolulu, HI, USA.} } \\
Copyright is held by the owner/author(s). Publication rights licensed to ACM. \\
ACM ISBN 978-1-4503-6708-0/20/04\ ...\$15.00.\\
http://dx.doi.org/10.1145/3313831.3376799}

\usepackage{balance}       
\usepackage{graphics}      
\usepackage[T1]{fontenc}   
\usepackage{txfonts}
\usepackage{mathptmx}
\usepackage{color}
\usepackage{booktabs}
\usepackage{textcomp}
\usepackage[dvipsnames]{xcolor}
\usepackage{colortbl}
\usepackage{array}
\usepackage{tabu}
\usepackage{mdframed}
\usepackage{framed}
\definecolor{shadecolor}{gray}{0.75}
\usepackage{tabularx}
\usepackage{float}
\usepackage{capt-of}
\usepackage[pdflang={en-US},pdftex]{hyperref}

\usepackage{microtype}        
\usepackage[all]{hypcap}    
\usepackage{ccicons}          

\def\plaintitle{TurkEyes: A Web-Based Toolbox for\\ Crowdsourcing Attention Data}
\def\plainkeywords{Eye tracking; attention; crowdsourcing; interaction techniques}

\makeatletter
\def\url@leostyle{%
  \@ifundefined{selectfont}{
    \def\UrlFont{\sf}
  }{
    \def\UrlFont{\small\bf\ttfamily}
  }}
\makeatother
\def\pprw{8.5in}
\def\pprh{11in}

\definecolor{linkColor}{RGB}{6,125,233}
\hypersetup{%
  pdftitle={\plaintitle},
  pdfauthor={\emptyauthor},
  pdfkeywords={\plainkeywords},
  pdfdisplaydoctitle=true, 
  bookmarksnumbered,
  pdfstartview={FitH},
  colorlinks,
  citecolor=black,
  filecolor=black,
  linkcolor=black,
  urlcolor=linkColor,
  breaklinks=true,
  hypertexnames=false
}

\begin{document}

\title{\plaintitle}

\numberofauthors{8}
\author{%
 Anelise Newman\textsuperscript{1},
 Barry McNamara\textsuperscript{1},
 Camilo Fosco\textsuperscript{1},
 Yun Bin Zhang\textsuperscript{2}, \\
 Pat Sukhum\textsuperscript{2},
 Matthew Tancik\textsuperscript{3},
 Nam Wook Kim\textsuperscript{4},
 Zoya Bylinskii\textsuperscript{5} \\
 \alignauthor{
   \affaddr{\textsuperscript{1}MIT}\\
   \email{\{apnewman, barryam3, camilolu\}@mit.edu}}
 \alignauthor{
   \affaddr{\textsuperscript{2}Harvard}\\
   \email{\{ybzhang, psukhum\}@g.harvard.edu}}
 \alignauthor{
   \affaddr{\textsuperscript{3}University of California, Berkeley}\\
   \email{tancik@berkeley.edu}}
 \alignauthor{
   \affaddr{\textsuperscript{4}Boston College}\\
   \email{nam.wook.kim@bc.edu}}
 \alignauthor{
   \affaddr{\textsuperscript{5}Adobe Inc.}\\
   \email{bylinski@adobe.com}}
}

\maketitle

\begin{abstract}
Eye movements provide insight into what parts of an image a viewer finds most salient, interesting, or relevant to the task at hand. 
Unfortunately, eye tracking data, a commonly-used proxy for attention, is cumbersome to collect. 
Here we explore an alternative: a comprehensive web-based toolbox for crowdsourcing visual attention. 
We draw from four main classes of attention-capturing methodologies in the literature. 
ZoomMaps is a novel \textit{zoom-based} interface that captures viewing on a mobile phone. 
CodeCharts is a \textit{self-reporting} methodology that records points of interest at precise viewing durations. 
ImportAnnots is an \textit{annotation} tool for selecting important image regions, and \textit{cursor-based} BubbleView lets viewers click to deblur a small area. 
We compare these methodologies using a common analysis framework in order to develop appropriate use cases for each interface.
This toolbox and our analyses provide a blueprint for how to gather attention data at scale without an eye tracker.
\end{abstract}

\ccsdesc[500]{Human-centered computing~Interaction techniques}
\ccsdesc[300]{Human-centered computing~Web-based interaction}
\ccsdesc[300]{Human-centered computing~Interactive systems and tools}
\ccsdesc[300]{Human-centered computing~User studies}
\ccsdesc[300]{Human-centered computing~Empirical studies in interaction design}

\keywords{\plainkeywords}

\printccsdesc

\setlength{\textfloatsep}{18pt}
\begin{figure}[h]
 \centering
\includegraphics[width=1\linewidth]{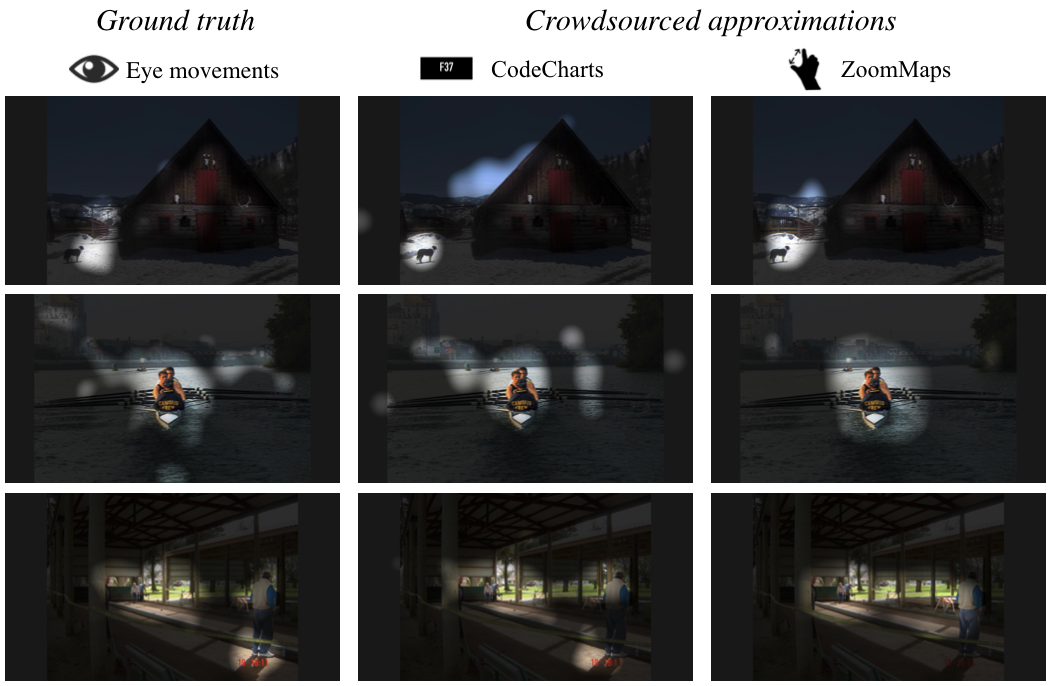}
 \caption{We consider approaches for crowdsourcing human visual attention data without the use of an eye tracker. 
 The attention heatmaps generated by two of our interfaces, CodeCharts and ZoomMaps, mimic heatmaps obtained using eye tracking. 
 While CodeCharts closely approximates eye movements, ZoomMaps gives a coarser approximation of attention with more emphasis on distant details.
 This paper will cover how the methodologies capture stable aspects of human attention and the unique features that make them suitable for different applications.
 }
 \label{fig:header}
\end{figure}

\section{Introduction}

Gaze provides a window into what aspects of an image, design, or visualization people find most engaging. 
Where someone looks on an image can predict whether they remember it or not \cite{beyondmem, bylinskii_intrinsic_extrinsic_2015}. 
Attention-grabbing regions of a poster can be used to summarize the design for later retrieval \cite{predimportance}. 
The most salient parts of an image can guide automatic cropping and retargeting \cite{twitter_engadget}. 
All of these applications rely on inferring where people are paying attention by capturing where they are looking. 

However, attention data has historically been difficult to collect at scale, as it involves in-lab eye tracking using dedicated hardware. 
The time it takes to recruit and run each participant prevents quick data collection and iteration. 
Meanwhile, online crowdsourcing allows for rapidly collecting large amounts of human data. 
Although webcam-based eye tracking has been proposed as a crowdsourceable alternative \cite{krafka2016eye,papoutsaki2016webgazer,zhang2018training}, it has many requirements, such as specific lighting conditions and participant pose, that are difficult to enforce. 
This has motivated a body of research on interactive user interfaces capable of capturing attention data without eye tracking.

In this paper, we analyze and expand the state-of-the-art in interaction methodologies for capturing attention.
We present TurkEyes (\url{http://turkeyes.mit.edu/}), a toolbox of four interfaces for gathering attention data using just a laptop or mobile phone.
None of the interfaces we consider explicitly measure eye movements. Rather, we make use of interaction methodologies from the literature that are correlated with visual attention (Fig. \ref{fig:header}). The interfaces we explore are:

\begin{tabular}{c m{6cm}}
\begin{minipage}{.08\textwidth}
  \includegraphics[width=\linewidth]{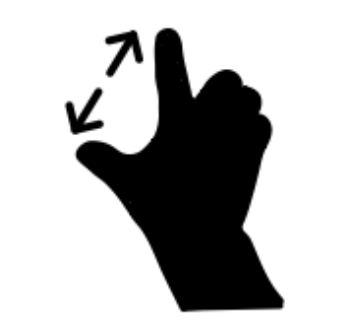}
\end{minipage} & \textbf{ZoomMaps (zoom-based):} Participants use the pinch-zoom gesture on a mobile phone to explore image content. \\
\begin{minipage}{.08\textwidth}
  \includegraphics[width=\linewidth]{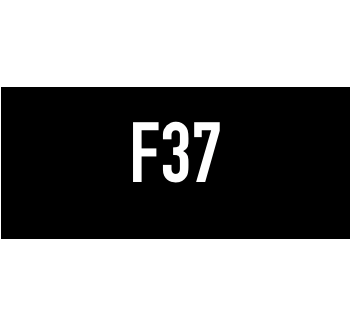}
\end{minipage} & \textbf{CodeCharts (self-report):} Participants specify where on an image they gazed using a grid of codes that appears after image presentation, inspired by \cite{rudoy2012crowdsourcing}.\\
\begin{minipage}{.08\textwidth}
  \includegraphics[width=\linewidth]{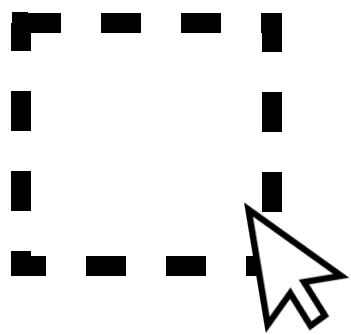}
\end{minipage} & \textbf{ImportAnnots (annotation):} Participants paint over regions of a design they consider important using binary masks \cite{odonovan}. \\
\begin{minipage}{.08\textwidth}
  \includegraphics[width=\linewidth]{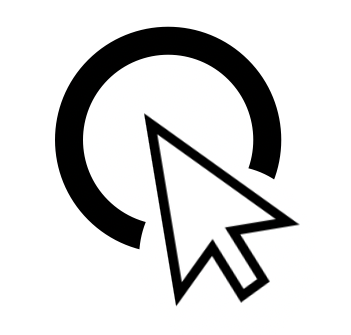}
\end{minipage} & \textbf{BubbleView (cursor-based):} Participants click to deblur/expose small bubble regions on an otherwise blurry image \cite{kim2017bubbleview}.  \\
\end{tabular}

\newcolumntype{P}[1]{>{\centering\arraybackslash}m{#1}}
\renewcommand{\arraystretch}{1.6}
\begin{table*}[t!]
\centering
\begin{tabular}{P{.9cm} P{.75cm}
    |P{4.6cm}|P{4.9cm}|P{4.8cm}
}
    \multicolumn{2}{c}{\textbf{Interface}} & \multicolumn{1}{c}{\textbf{Use Case}} & \multicolumn{1}{c}{\textbf{Advantages}} & \multicolumn{1}{c}{\textbf{Drawbacks}} \\
     \hline
     \textbf{Zoom \newline Maps} & \begin{minipage}{.02\textwidth}
        \hspace*{-0.25cm}
        \includegraphics[width=2\linewidth]{figures/icon-zoommaps.png}
    \end{minipage} & Capturing exploration of large images at multiple scales & Works on images with multi-scale content, natural form of interaction & Coarse approximation of attention \\ 
     \textbf{Code \newline Charts} & \begin{minipage}{.02\textwidth}
        \hspace*{-0.15cm}
        \includegraphics[width=2\linewidth]{figures/icon-codecharts.png}
    \end{minipage} & Approximating eye-tracking, esp. for precise viewing durations & Doesn't distort stimuli, experimenter controls timing, fun & Little data per participant, images must fit on screen \\
     \textbf{Import \newline Annots} & \begin{minipage}{.02\textwidth}
        \hspace*{-0.15cm}
        \includegraphics[width=2\linewidth]{figures/icon-importannots.png}
    \end{minipage} & Comparing importance of graphic design elements & Produces clean segmentations, captures importance & Not ideal for natural images, measures importance over attention \\
     \textbf{Bubble \newline View} & \begin{minipage}{.02\textwidth}
        \hspace*{-0.15cm}
        \includegraphics[width=2\linewidth]{figures/icon-bubbleview.png}
    \end{minipage} & Approximating eye-tracking, esp. during description tasks & Versatile, cheap & Distorts stimuli and viewing experience \\
    \hline
\end{tabular}
\caption{Use cases and trade-offs for the four TurkEyes interfaces.}
\label{tab:advantages}
\end{table*}

We design our toolbox to represent four main categories of attention-capturing interfaces that we identify in the literature.
Two of these categories lacked a well-studied concrete implementation, so we created novel interfaces (ZoomMaps and CodeCharts) to fill this gap.
For the other two categories we draw on existing tools (ImportAnnots and BubbleView).
We integrate these interfaces into a shared software framework and provide code to convert their attention data into a common format so that they can be directly compared. 

Next, we do a deep-dive on these interfaces by conducting extensive experiments on Amazon's Mechanical Turk. 
We carefully design tasks and validation procedures to produce high-quality data.
We collect data on a variety of stimuli (natural and non-natural images) to determine what insights are discoverable by each tool. 
Although all the interfaces capture some common aspects of attention, they are best suited for different image types and tasks, and we provide guidelines as to applicable use cases for each.

Our contributions are:
1) a comprehensive toolbox that gathers attention-gathering interfaces into a common code and analysis framework and 
2) a user guide explaining how, when, and why to deploy each interface to gather attention data tailored to a particular use case. 

\section{Related Work}

Interaction data such as mouse/keyboard on desktop and touch/zoom on mobile provides a window into what people find relevant and interesting in online content \cite{chen2001can, huang2012user, rodden2008eye, xu_attention_in_guis_2016}.
However, these same interaction methodologies can be harnessed to capture attention on images as a replacement for in-lab eye tracking. 

\textbf{Eye tracking.} Eye movements collected using dedicated hardware have long been used to quantify attention. 
Researchers have also used built-in webcams to obtain coarse-grained attention data from crowdworkers~\cite{krafka2016eye,papoutsaki2016webgazer}, but these methods are insufficiently robust, requiring controlled conditions. 
Efforts have thus turned to interaction techniques that approximate eye movements, falling into one of the following four categories. 

\textbf{Cursor-based interfaces.} Prior work investigated the correlation between mouse and gaze locations~\cite{guo2010towards,huang2012user,rodden2008eye}. Cursor movements can complement eye movements, especially when a participant can use both to interact with visual content. A separate line of work considered cursor-based interfaces as a proxy for eye tracking~\cite{bednarik2005effects,jiang2015salicon,schulte2011flashlight}. For instance, the moving-window methodology reveals only portions of an otherwise-obscured image depending on where a user positions the mouse cursor~\cite{jansen2003tool,mcconkie1975span,rayner2014gaze,tatler2005visual}. 
This is the basis of the BubbleView methodology, which was extensively explored in~\cite{kim2017bubbleview} and provides a well-understood comparison point for our work.

\textbf{Self-report interfaces.} Moving-window methodologies like BubbleView distort the underlying image. An alternative is to show viewers an undistorted image and ask them to report where they looked, often with the aid of an annotated grid \cite{cheng_gazecrowd_2015, rudoy2012crowdsourcing}. Our CodeCharts interface is based on \cite{rudoy2012crowdsourcing}. 

\textbf{Zoom-based interfaces.} 
Zoom allows users to expand content that they find engaging and want to view in greater detail \cite{bartram_fisheye_1995,bier_magiclens_1993}. 
Previous work investigated the zoomable viewport on a mobile phone as a measure of user engagement with an interface or list of search results \cite{
guo_mining_touch_2013,
guo_large_scale_analysis_2016, 
lagun_towards_better_measurement_2014, 
lagun_understanding_user_attention_2016, 
lamberti_supporting_web_analytics_2017,
li_towards_measuring_2017}. Huang et al. even proposed generating heatmaps based on the viewport \cite{Huang2012WebUI}.
Our ZoomMaps methodology expands on this work by using viewport data to produce an attention heatmap on an arbitrary image, treating the mobile phone as a restricted window through which users explore areas of interest.

\textbf{Annotation interfaces.} UI tools for collecting object segmentations in images were developed to produce training data for computer vision tasks such as object detection and recognition \cite{russell2008labelme, salvador_ask_n_seek_2013}. However, they have also been used to identify graphic design elements that a viewer rates as important. ImportAnnots refers to the interface for capturing explicit ``Importance Annotations'', first introduced by O'Donovan et al.~\cite{odonovan}, and has been used to collect data for training computational models to predict importance of graphic designs~\cite{predimportance,odonovan}. 

\section{Considerations for evaluating attention}

Here we discuss ideas, tools, and analysis methods for evaluating attention interfaces. 
These considerations will motivate an in-depth look at each interface and guide our comparisons between them. 

\textbf{Four classes of interfaces.} We group previous work on attention-gathering interfaces into four categories. 
\textit{Zoom-based} interfaces use a viewer's zoom patterns as a signal of interest in regions that are viewed or zoomed. 
\textit{Self-report} interfaces show an image for a limited time and ask the viewer to report where they were looking using a visual guide.
\textit{Annotation} interfaces allow users to explicitly segment regions they judge to be relevant to the task at hand.
\textit{Cursor-based} interfaces leverage correlations between mouse movements and eye movements, often by incentivizing the viewer to click/hover to explore points of interest.
Our toolbox contains one exemplar of each and our analyses will consider the capabilities and drawbacks of these approaches.

\hypertarget{metrics_anchor}{
\textbf{Representations and metrics for comparing attention.}}
We convert the output from all of our interfaces into a common representation: an \textit{attention heatmap}, where regions with higher heatmap values are more attended to. 
This is significant because it lets us directly compare output from all the interfaces. 

To quantify the similarity of attention data captured in different ways, we use the Pearson's Correlation Coefficient (CC) and Normalized Scanpath Saliency (NSS) metrics.
CC and NSS are the preferred metrics for evaluating saliency predictions and are highly correlated \cite{salMetrics_Bylinskii}. 
CC measures the pixel-wise correlation between two normalized heatmaps and ranges from -1 (inversely correlated) to 1 (perfectly correlated); we use it to compare attention heatmaps generated by different interfaces.
NSS measures the mean value of a normalized attention heatmap evaluated at ground-truth eye fixation locations and ranges from 0 (no heatmap density at fixated locations) to infinity (all heatmap density at fixated locations). 
We report NSS when comparing our attention heatmaps to ground truth eye fixations, which does not require post-processing the fixations into a heatmap (in contrast to CC). 
These metrics were used in \cite{kim2017bubbleview} and studied in detail in \cite{salMetrics_Bylinskii}. \label{paragraph:metrics-explanation}

\textbf{Types of stimuli.} To investigate how well our interfaces function for different image types, we collect data using a subset of our interfaces on natural images, resumes, graphic designs, infographics, and data visualizations. The natural images are drawn from the CAT2000 dataset~\cite{borji2015cat2000} which has ground-truth eye tracking data. For 35 images from CAT2000, we collected data using all four interfaces: ZoomMaps, CodeCharts (at 6 different viewing durations), ImportAnnots, and BubbleView. We also evaluated our interfaces on 116 resumes and 20 graphic designs that we downloaded from Canva.com. 
Additionally, we ran ZoomMaps on larger, more complex images than the other interfaces: infographics from MASSVIS~\cite{beyondmem} and data visualizations from personal collections. 

\textbf{Types of task.} 
There are several common viewing tasks used when collecting attention data, including: search (looking for a particular element in an image), description (describing or annotating an image \textit{while} viewing it), memory (recalling some aspect of an image \textit{after} viewing it), and free-viewing (exploring the image freely with no explicit task).

A limitation of relying on interaction to collect attention is that the interface cannot be completely decoupled from the task.
For instance, it is impossible to have a free viewing ImportAnnots task or a description CodeCharts task. 
For crowdsourcing, it also important for (1) the task to incentivize participants to engage with the interface, and (2) data quality to be easily validated. 
For our experiments, we choose tasks that align with both the interaction methodology of each interface and the incentives of crowdworkers, and we explain how to validate the quality of the data captured with our procedures. 
For BubbleView and CodeCharts, we use a free-viewing task because the actions of clicking and reporting codes, respectively, naturally encourage interaction.
For ZoomMaps we use a memory task to encourage participants to explore the image by zooming in on details. 
For ImportAnnots, we use the annotation (description) task intrinsic to the interface, and we also report on the description task using BubbleView.

\textbf{Evaluation criteria.} 
We will consider several criteria when evaluating these interfaces, including: cost of data collection per image, type of stimuli and task that is appropriate for each interface, similarity of the data collected to eye movements, and what exactly each interface is measuring.
Table \ref{tab:advantages} provides a high-level summary of the benefits and drawbacks of each. 
We present this table up front in order to contextualize the discussion of the details of each interface in the next section.

\section{Introducing the TurkEyes Toolbox} 

In this section, we do a deep dive into the individual interfaces to describe how they work and present our experimental procedures for collecting attention data with each. 

\begin{tabular}{m{1.5cm} c}
\subsection{ZoomMaps} &
\begin{minipage}{.04\textwidth}
  \includegraphics[width=\linewidth]{figures/icon-zoommaps.png}
\end{minipage}
\end{tabular}

The mobile screen provides a naturally restricted window that is frequently used to explore multi-scale content with the help of the zoom functionality. 
We build a novel interface to capture the zoom patterns of participants viewing images on their mobile phones and show that these patterns can be used as an approximation of visual attention. 

\begin{figure}
 \centering
\includegraphics[width=0.7\linewidth]{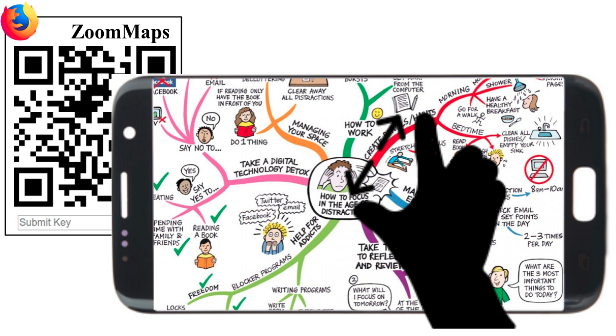}
 \caption{ZoomMaps UI. Participants use the pinch-zoom gesture on their phones to explore image content at higher resolutions.}
 \label{fig:zoommaps_UI}
\end{figure}

\textbf{Task flow.} Participants are sent to a landing page that contains a QR code and a URL that they use to open an image gallery in their mobile browser. 
They are instructed to spend a minimum amount of time (5-15 seconds per image depending on the experiment) exploring each image by panning and zooming. Depending on the experiment, they either answer questions about each image (in which case they answer questions on mobile) or fill out a task-specific questionnaire at the end just before submission (which can be completed on a desktop). 
Once they are done viewing the images on mobile, participants receive a completion code to enter on the landing page and receive credit for the task.

\textbf{Implementation.} We built an image gallery webpage augmented with tracking capabilities using the PhotoSwipe JavaScript library\footnote{\url{https://github.com/dimsemenov/photoswipe}}. We modify the library to capture any changes to the visible region of the image along with a timestamp. The interface allows pinching to zoom in or out on an image and swiping to switch images (Fig.~\ref{fig:zoommaps_UI}). The interaction data contains viewport coordinates on the image and a timestamp for every \emph{event} triggered by the user (i.e., the image is re-scaled or re-positioned on the screen). 

\textbf{Validation procedure.} 
We require that participants spend at least 1-5 seconds (depending on the experiment) on 85\% of the images and at least 3 minutes in total on a task that we estimate should take 4-9 minutes. 
Furthermore, we require that participants zoom on at least 20\% of the images. 
These thresholds were chosen empirically based on pilots. 
In most image collections, there are at least a few images that are uninteresting, receiving little viewing time, and there are frequently images that require no zooming because all elements are visible at the original scale.
Participants are encouraged to keep exploring until they have met our time or zoom requirements; this ensures that we do not need to discard data retroactively.

\begin{figure}[t]
 \centering
\includegraphics[width=0.75\linewidth]{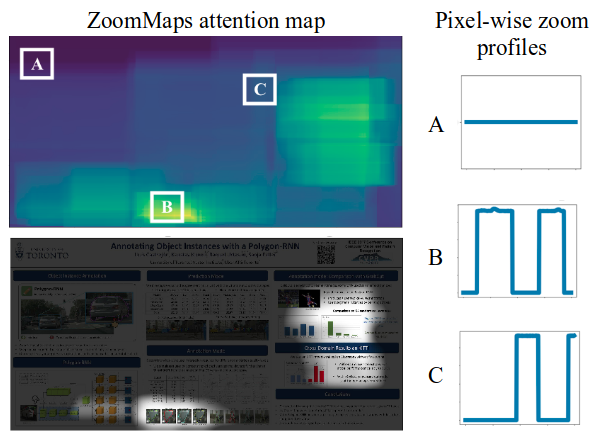}
 \caption{A ZoomMaps attention heatmap, visualized alone (top) and overlaid on the image used to create it (bottom). For three pixels, labeled A, B, and C, zoom over time is averaged to produce a value in the corresponding attention heatmap.}
 \label{fig:zoomlevel}
\end{figure}

\textbf{Generating attention heatmaps.} Our mobile interface stores the bounding boxes of zoomed image regions along with timestamps of when they were in focus. We use this information to extract which parts of the image were viewed for how long and at what zoom level. We then construct the attention heatmap as follows: for every pixel in the image, we compute its average \emph{zoom level} over the entire viewing interval. We define the zoom level for an image region as the full image area divided by the area of the image region that has been magnified. We assign this zoom level to all the pixels contained in the image region. We then compute each pixel's average zoom level over the viewing duration to obtain a ZoomMaps attention heatmap (Fig.~\ref{fig:zoomlevel}). Higher values in the heatmap correspond to regions of the image that were inspected with closer zoom on average.

\textbf{Cost.}
Participants were paid \$1.00-\$1.25 to look at 5-35 images. 
Payment depended on how many questions we asked and how long participants should spend exploring each image, where we assumed that more complex/larger visuals like infographics would take longer to explore than natural images. 
For twenty impressions per image, data collection cost \$0.72 per natural image, \$1.33 per resume, and \$5 per infographic. 

\textbf{Likeability.} Participants generally enjoyed the task and transitioned smoothly from browser to mobile. They were occasionally frustrated when asked to spend more time exploring the image after failing to meet our time or zoom requirements.

\begin{tabular}{m{1.7cm} c}
\subsection{CodeCharts} &
\begin{minipage}{.05\textwidth}
  \includegraphics[width=\linewidth]{figures/icon-codecharts.png}
\end{minipage}
\end{tabular}

\begin{figure}[b]
 \centering
\includegraphics[width=0.8\linewidth]{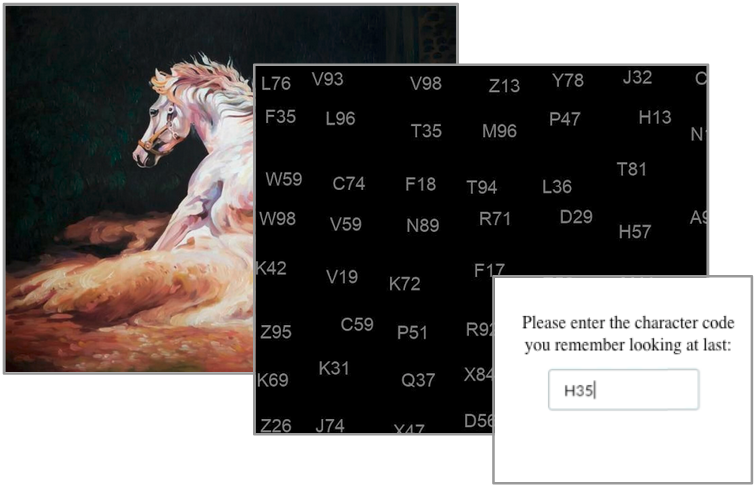}
 \caption{CodeCharts UI. Participants self-report a region of an image they gazed at using a grid of codes that appears after image presentation.}
 \label{fig:codecharts_UI}
\end{figure}

CodeCharts collects individual gaze points by asking participants to self-report where they were looking on an image using a grid of three-character codes called a \textit{codechart}. Because the timing of image presentation is controlled by the experimenter, it is possible to collect attention data at precise viewing durations. 

\textbf{Task flow.} A participant views an image on their screen for a preset duration, typically a few seconds. When the image disappears it is replaced by a jittered grid of three-character codes. The participant notes the last triplet they see when the image vanishes, which approximates where on the image they were looking at the time. When the codechart vanishes they self-report this alphanumeric code. This process, visualized in Fig. \ref{fig:codecharts_UI}, repeats for a sequence of images. Trials are separated by a fixation cross to re-center the participant's gaze.

\textbf{Implementation.} Each codechart is a grid of three-character alphanumeric codes. Each triplet is placed with a slight random horizontal and vertical jitter. Every time an image is presented it is accompanied by a different codechart. We pad and resize all input images to a consistent size and generate codecharts with the same dimensions so they can be easily displayed in the browser. Both target images and codecharts are resized to fit into a display window of 1000 by 700 pixels and triplets are displayed at a font size of around 16 pixels. 
To fit this display size, our CodeCharts implementation was designed for desktop, but the same methodology could be adapted for mobile.

\begin{figure}
 \centering
\includegraphics[width=0.99\linewidth]{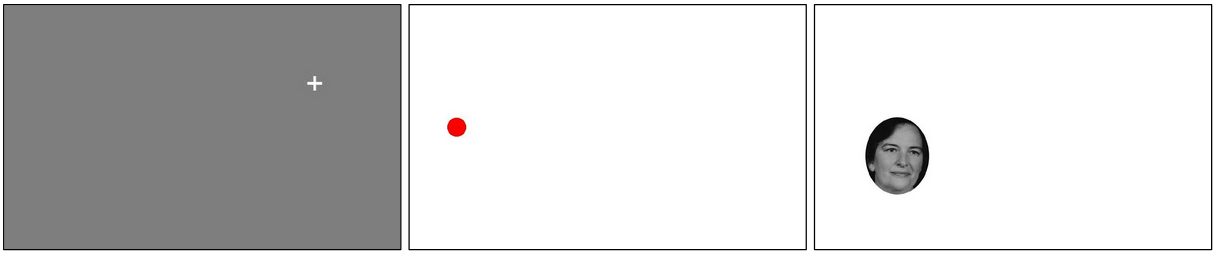}
 \caption{Sample validation images for the CodeCharts UI. We experimented with a regular fixation cross (left) and a red circle on a white background (center) before selecting a cropped face on a white background (right) as the image that most effectively encouraged participants to move their eyes to the cue. 
 Participants were expected to type a triplet code that overlapped with the coordinates of the cue. 
 The cues are plotted larger than actual scale for ease of viewing.}
 \label{fig:sentinels}
\end{figure}

The goal is to display the codechart long enough for participants to read a triplet but not so long that their eyes can wander. We found that 400ms was the optimal exposure duration. By systematically decreasing codechart exposure from 750ms to 400ms, the NSS similarity of CodeCharts data to ground truth eye movements increased from 1.74 to 1.89. When we dropped the exposure time to 300ms, we saw a significant drop (24\%) in accuracy at reporting a valid code. 

One limitation of the interface is that the codechart can produce gridlike artifacts in the output. This can be mitigated by jittering the axes of the whole grid instead of jittering individual codes within a rectangular cell. Experimenters must take care that triplets are well-spaced and extend to the edge of the image to avoid collecting skewed data.

\textbf{Validation procedure.} Some of the images shown in our task are validation images. These images have a plain background and a single point of interest that aligns with one or more ``correct'' triplets in the corresponding codechart. We tried three different styles of validation images: a plain fixation cross, a red circle on a white background, and a cropped face image on a white background (Fig.~\ref{fig:sentinels}). 
Face images were taken from the face dataset compiled by Bainbridge et al. \cite{  Bainbridge2013_faces}. We found that faces provide a more interesting cue than simpler stimuli, incentivizing participants to attend to the cue. We also found that explicitly instructing participants to look at validation images increased similarity to ground-truth eye movements over all experiment images (not just the validation ones!).

Our task starts with a screening phase of three normal and three validation images where participants must correctly enter all validation codes and may only enter one nonexistent code that does not appear on the codechart.
Validation images are also interspersed throughout the sequence and are used to retroactively discard data from participants who miss over 25\% of validation codes. 
We also eliminate participants who look at the same spot on the screen for many images in a row. 
We find that participants with higher validation accuracy produce data that is more similar to human eye movements, which justifies our attempts to select interesting validation images. 
This could be because these participants are more attentive or get used to moving their eyes in response to stimuli. 

\textbf{Generating attention heatmaps.} We combine all the gaze points for one image (one per participant) and blur them with a sigma of $50$ to generate a heatmap. Our triplets are spaced approximately 100 pixels apart, so 50 is a good approximation of the radius of uncertainty in the interface.

\textbf{Cost.} 
For a 48-trial experiment on natural images, participants spent on average 1 minute reading instructions and just over 6 minutes on the rest of the task including a demographic survey. At an hourly rate of \$10, this works out to \$1.25-\$2.00 per image for 50 participants' worth of data, depending on how long the images were shown.

\textbf{Likeability.} The task was very well-received. Participants often described it as ``fun'' and ``interesting'' while also being ``hard'' and ``fast''. We think that the automatic timing of the task contributed to a game-like experience. 

\begin{tabular}{m{2cm} c}
\subsection{ImportAnnots} &
\begin{minipage}{.04\textwidth}
  \includegraphics[width=\linewidth]{figures/icon-importannots.png}
\end{minipage}
\end{tabular}

\begin{figure}
 \centering
\includegraphics[width=0.7\linewidth]{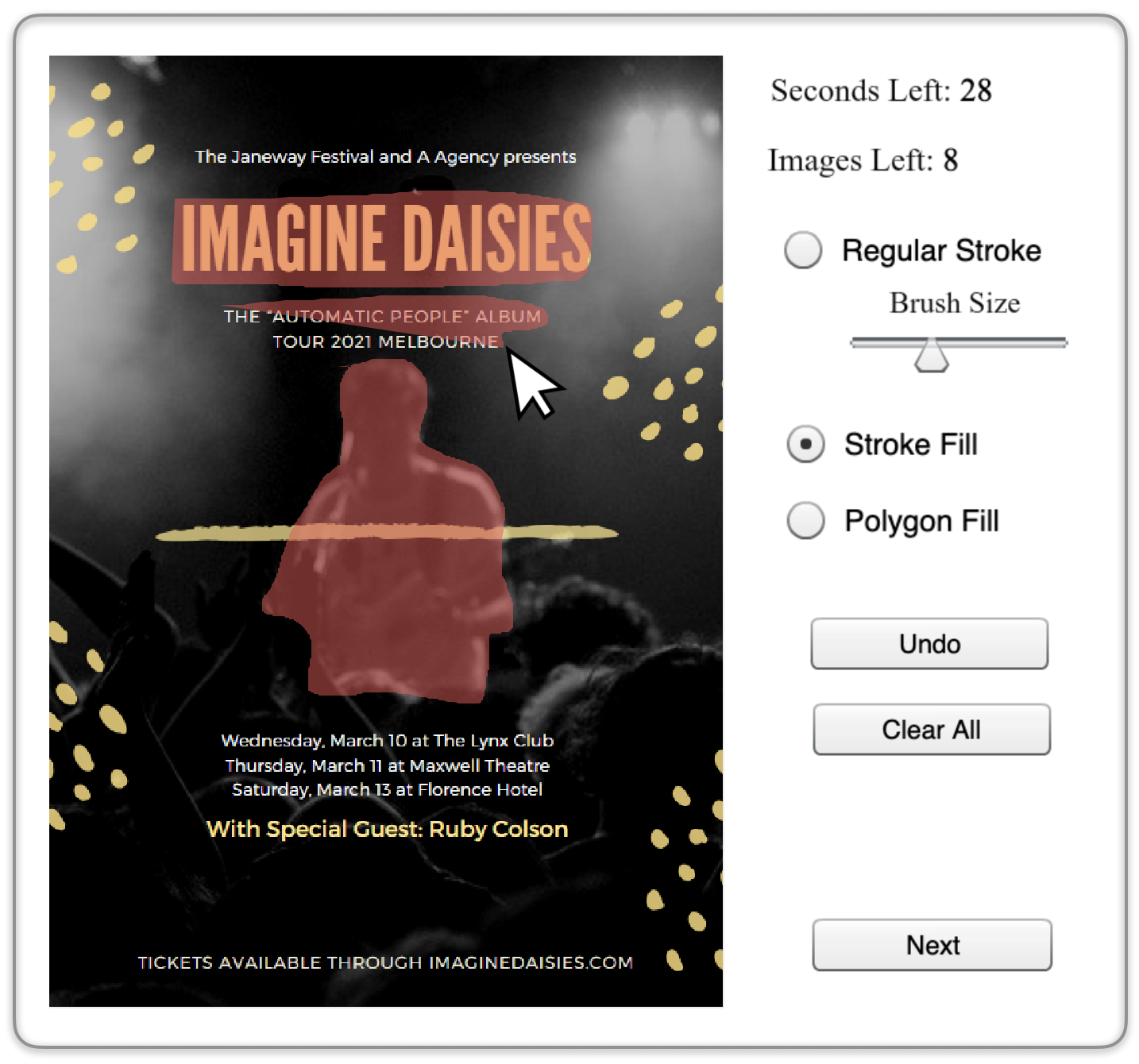}
 \caption{ImportAnnots UI. Participants paint over regions of a design that they consider important using binary masks. The mask is shown as a transparent red overlay.}
 \label{fig:importannots_UI}
\end{figure}

O'Donovan et al. introduced the idea of having crowdworkers annotate important elements on graphic designs using binary masks and averaging them to construct importance heatmaps \cite{odonovan}. 
For this paper, we re-purpose the initial interface, add a validation procedure, and test the interface on different image types including natural scenes, infographics, and resumes.

\textbf{Task flow.} Participants are presented with a series of images one at a time and are asked to annotate the most important regions (Fig.~\ref{fig:importannots_UI}). There are no definitions of what should be considered ``important''. We restrict participants to a maximum of one minute per image.

\textbf{Implementation.} We used legacy code from O'Donovan et al.~\cite{odonovan} for the annotation tool embedded in our task interface. 
It is a Flash application, designed for desktop, that provides 3 annotation tools: \emph{stroke fill} that allows tracing the contours of an object to provide a fine-grained segmentation, \emph{polygon fill} that allows plotting points with connected lines for coarser annotation, and \emph{regular stroke} for painting over a region. 
Stroke fill is set as the default and is what most participants choose: 64\% of images are annotated using stroke fill compared to 34.2\% with polygon fill and 1.8\% with regular stroke.

\begin{figure}
 \centering
 \includegraphics[width=0.99\linewidth]{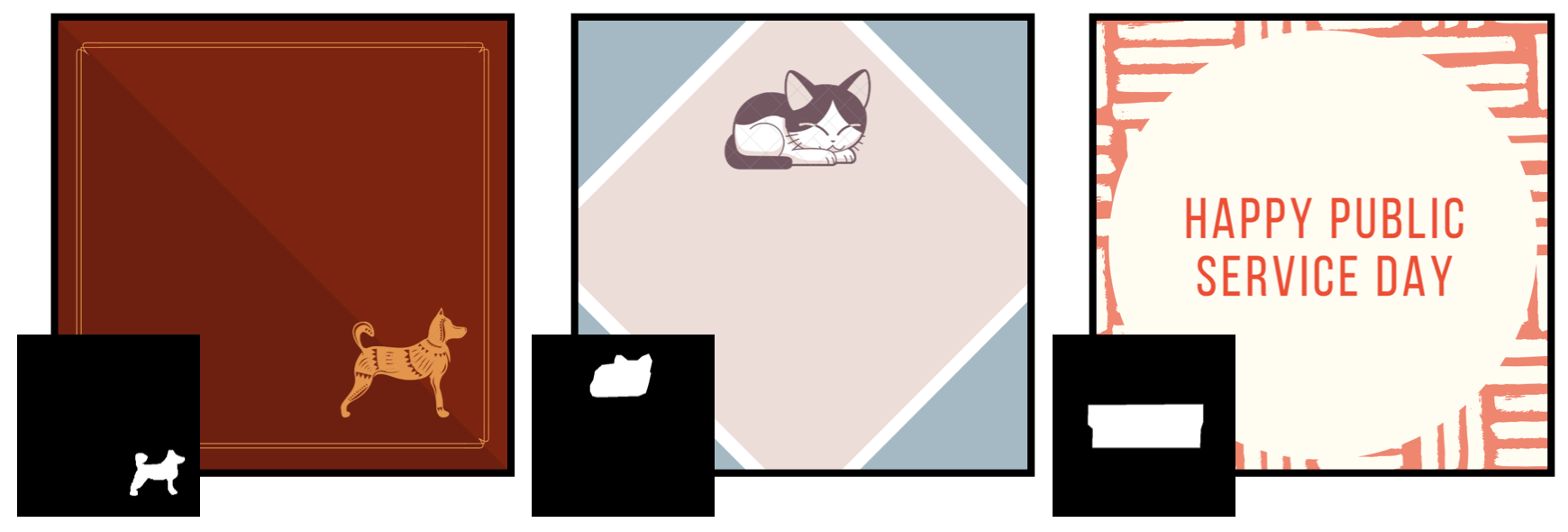}
 \caption{Sample validation images for the ImportAnnots UI: we simplified vector graphic designs to contain a single element that unambiguously stood out. Throughout the task, we computed intersection-over-union of participant annotations with ground truth annotations (insets) to ensure annotation quality.}
 \label{fig:importannots-sentinels}
\end{figure}

\textbf{Validation procedure.} Interspersed throughout the task were validation images in the form of graphic designs containing one main textual or graphical element (Fig.~\ref{fig:importannots-sentinels}). Forty such validation images were constructed by manually deleting extra elements from existing vector designs to make annotation of importance more obvious. 
For a 5-minute task, participants annotated 10 images and 3 validation designs in a random order. To meet our quality thresholds, participants needed to annotate at least one object in all but one of the images and to correctly annotate 2/3 of the validation designs. Correctness on the validation designs was measured by an intersection-over-union (IoU) threshold of 0.55, where IoU is the area of overlap of the validation element and the user's selection over the total area of the two. 
We set this threshold by running a pilot experiment to capture reasonable variation in validation annotations and then calculating the mean IoU across annotations manually selected to be of high quality.
We only collected data from participants who met the validation threshold and did not discard data retroactively.

\textbf{Generating attention heatmaps.} Each participant generates one binary mask per image. The binary masks are averaged across participants to produce an overall attention heatmap for the image. Despite high inter-observer variability and noisy annotations, averaged over a large number of participants (20-30), the mean importance maps give a plausible ranking of importance (see appendix of \cite{odonovan}).

\textbf{Cost.} We paid participants \$0.85-\$1.00 for annotating 10 designs, bringing total costs to \$2.55-\$3.00 per image for 30 annotations.

\textbf{Likeability.} Some of the participants found that the tools were not immediately intuitive. In general, this task takes longer per image than others because of the level of annotation required.

\begin{figure}[H]
 \centering
\includegraphics[width=0.9\linewidth]{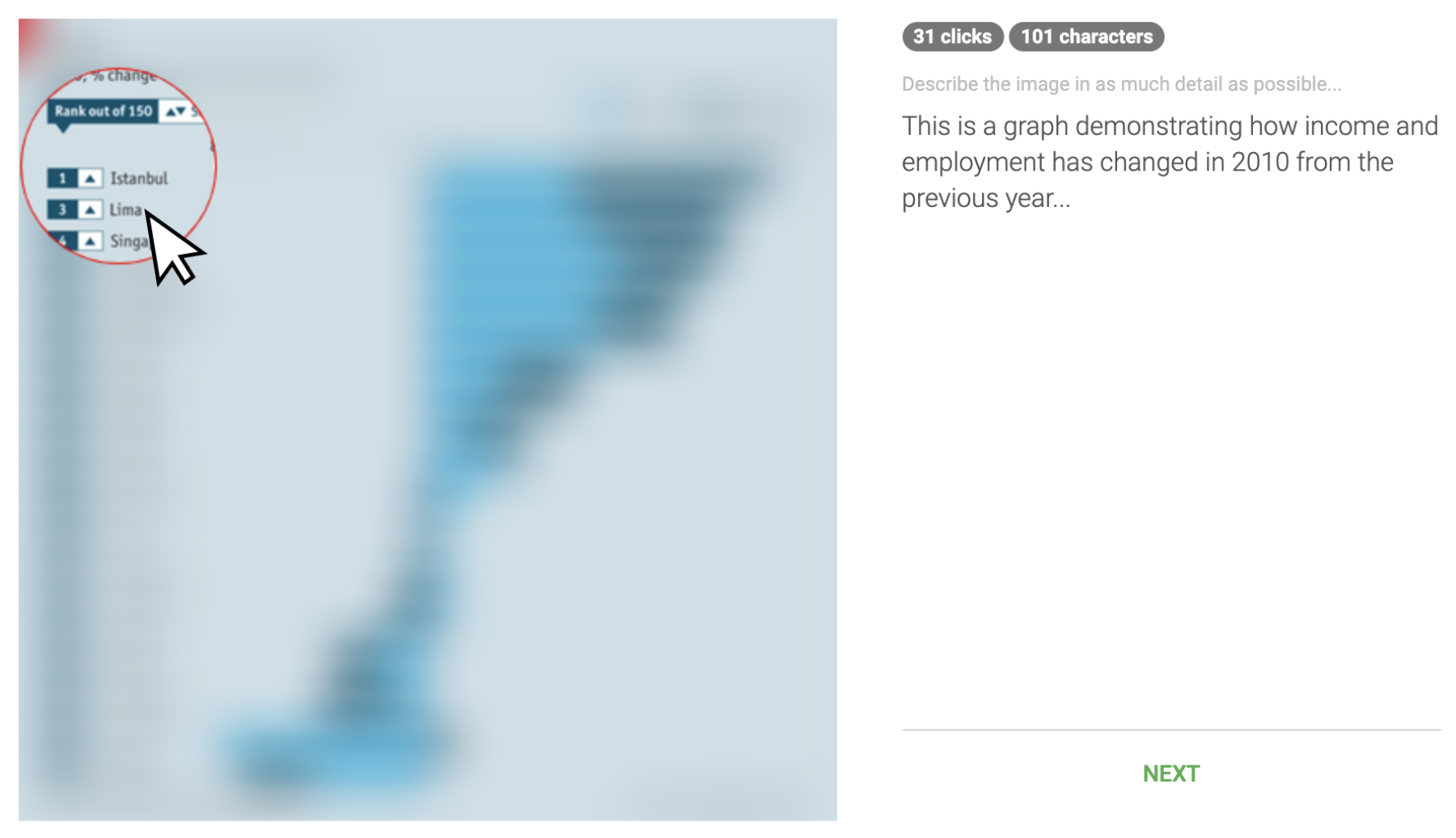}
 \caption{BubbleView UI. Participants click to deblur/expose small regions of an otherwise blurry image.}
 \label{fig:bubbleview_UI}
\end{figure}

\begin{tabular}{m{1.5cm} c}
\subsection{BubbleView} &
\begin{minipage}{.04\textwidth}
  \includegraphics[width=\linewidth]{figures/icon-bubbleview.png}
\end{minipage}
\end{tabular}

BubbleView is a cursor-based moving-window methodology. The original image is blurred to distort text regions and disable legibility, requiring participants to click around to de-blur small, circular ``bubble'' regions at full resolution (Fig.~\ref{fig:bubbleview_UI}). 
BubbleView was initially introduced in~\cite{kim2015crowdsourced}, which included a thorough comparison to eye movements on different image types and tasks. We reuse some of those analyses in this paper.

\textbf{Task flow.} Participants are asked to explore one image at a time using their mouse cursor to click and deblur regions of an image. 
Clicking on a new location re-blurs the previously clicked location. 
Blurring the image loosely approximates peripheral vision.
Kim et al. found that a blur radius of 30-50 pixels, corresponding to 1-2 degrees of visual angle, produces clicks that most closely approximate eye movements, without hindering search~\cite{kim2017bubbleview}.
For free-viewing tasks (no instructions other than to freely explore the image), viewing time is fixed, while for description tasks, participants are presented with a text box on the side of the image where they are asked to describe the image. As noted in Kim et al.~\cite{kim2017bubbleview}, it takes 2-3 times longer for participants to explore an image with BubbleView than to view it naturally for the same number of gaze points in the same unit time. In other words, this interface slows down visual processing relative to natural viewing. 

\textbf{Implementation.} BubbleView exposes several parameters to the experimenter: the blur sigma used to distort the image, radius of the bubble region exposed on-click, task timing and set-up (whether or not to include a required text field entry with each image), and whether discrete mouse clicks or continuous mouse movements are collected. Different task types warrant different parameter choices to best approximate ground truth eye movements; we refer to Kim et al.~\cite{kim2017bubbleview} for the details.

\textbf{Validation procedure.} At task-time the only validation procedure was a minimum number of 150 characters required in the text entry box for describing each image. 
During post-processing, participant data that included too few clicks was removed. These thresholds were set empirically (to 10 clicks/image for a description task and 2 clicks/image for a free-viewing task~\cite{kim2017bubbleview}). Interquartile range-based outlier computation~\cite{komarov2013crowdsourcing} was also used to remove too few or too many clicks as a secondary threshold.
This eliminated on average 2\% of participants during postprocessing \cite{kim2017bubbleview}.

\textbf{Generating attention heatmaps.} Given a set of BubbleView mouse clicks on an image, an attention heatmap is computed by blurring the click locations with a Gaussian with a particular sigma (a different one per image dataset~\cite{kim2017bubbleview}).

\textbf{Cost.} Adjusting BubbleView's hourly rate from \$6 to \$10 (to be comparable to the compensation we use in our tasks), and accounting for data discarded by quality checks, we can re-estimate the numbers in Table 7 of~\cite{kim2017bubbleview}: \$0.45 per image for 15 participants and a free-viewing duration of 10 seconds, including the cost of discarded data.

\textbf{Likeability.} Participants' experience with this interface depended in part on the experimental parameters used. Kim et al.~\cite{kim2017bubbleview} indicated that participants complained about the difficulty and tediousness of the task at small bubble sizes. 

\section{Choosing a tool}

In this section, we analyze the data we collected with our interfaces to evaluate them along various axes of interest. 
The interaction methods of each interface make each suitable to different types of tasks and stimuli. 
Additionally, the attention heatmaps generated by each method differ in some significant ways; we consider why and the implications for researchers. 

\subsection{How many participants are required?} 

\begin{figure}
 \centering
\includegraphics[width=1\linewidth]{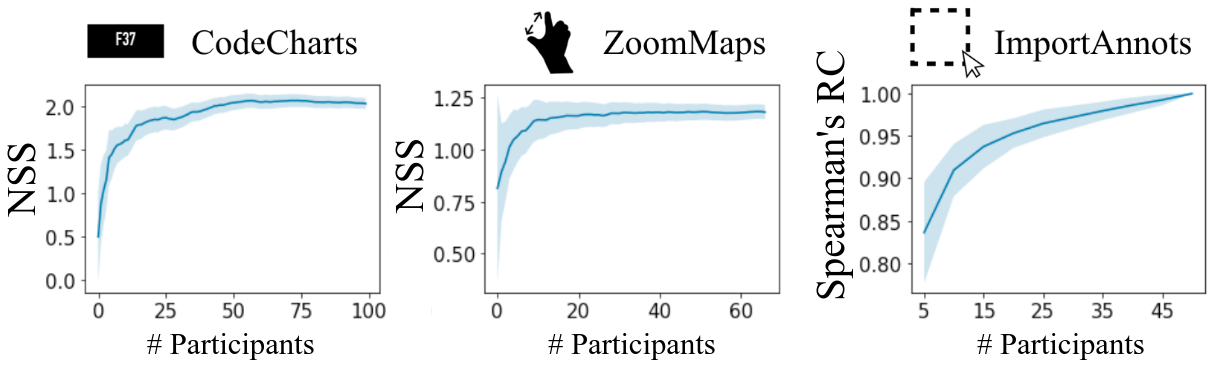}
 \caption{Performance improves with number of participants for CodeCharts, ZoomMaps, and ImportAnnots. We chose our recommended number of participants to achieve 98\% of the performance with the maximum number of participants tested. 
 For CodeCharts and ZoomMaps, we use NSS similarity to eye movements to measure performance. 
 For ImportAnnots, which is more tailored to ranking graphic design elements than approximating eye movements, we use Spearman's rank correlation over ranked graphic design elements, where we compare the ranking produced by a limited number of participants to that produced by many participants. 
 }
 \label{fig:num-participants}
\end{figure}

The interfaces collect different amounts of data per participant.
For instance, CodeCharts yields just a single gaze point, whereas BubbleView produces many clicks per participant. 
Thus, we calculate for each interface how many participants we need to obtain a stable measure of attention on an image. 
For BubbleView, CodeCharts, and ZoomMaps, we measure performance as the NSS similarity of the generated attention heatmaps to ground-truth eye fixations.
Similar to Kim et al.~\cite{kim2017bubbleview}, we calculate the number of participants that yields 98\% of maximum performance for a given interface. Our results are in Fig. \ref{fig:num-participants}. 
Kim et al. report that after about 10-15 participants, the NSS similarity of BubbleView was already 97-98\% of the performance achievable with many more participants~\cite{kim2017bubbleview}. 
For the rest of our interfaces, we use performance with all participants as an upper bound. 
We find that CodeCharts requires 50 participants per image, significantly more than BubbleView. 
ZoomMaps requires 15-20 participants. 
For ImportAnnots, direct comparison to eye tracking data is not as meaningful because by design, the generated heatmaps have a different structure than eye tracking data (see the low similarity in Table \ref{tab:eye-similarity}). 
As such, we use a performance metric more tailored to evaluating the importance of graphic design elements. We use the ImportAnnots heatmap to assign an importance score per element, taking the maximum value of the heatmap across an element, as in~\cite{predimportance}. We then use these importance scores to rank the elements in a graphic design.
We find that the ImportAnnots rankings produced by 30 participants are closely aligned (Spearman's rank correlation = 0.98) with those obtained using all participants. 

\subsection{How much will it cost?} 

\begin{figure}
 \centering
\includegraphics[width=0.7\linewidth]{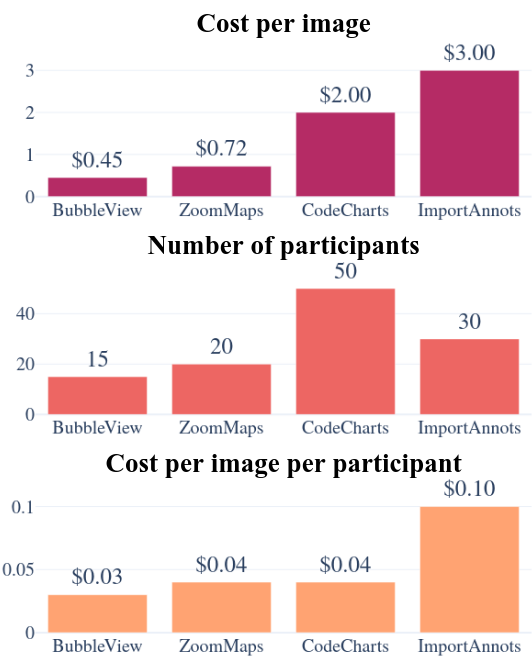}
 \caption{
 Cost comparison (on natural images). We find that differences in price per image for each interface (top row) are driven more by number of participants required (middle row) than differences in the price of one person attending to an image (bottom row). ImportAnnots is a special case because in addition to reporting attention, it requires the participant to segment objects, which drives up the time required per image and its price.}
 \label{fig:costs}
\end{figure}

The cost of obtaining an attention heatmap for an image is (number of participants) $\times$ (cost per image per participant). 
The results of this calculation for our experiments on natural images are shown in the first row of Fig. \ref{fig:costs}.
BubbleView is the cheapest at \$0.45 per image, followed by ZoomMaps, CodeCharts, and ImportAnnots at \$3.00. 

What drives this difference in the cost of attention? 
We observe that across interfaces, there is a remarkable consistency in cost per image per participant (bottom row of Fig. \ref{fig:costs}). 
We paid on average 2-4 cents per image per participant in the CodeCharts experiments, depending on how long the image was shown. 
The per-image per-participant costs of BubbleView and ZoomMaps are around 3 and 4 cents, respectively.
As a comparison point, for in-lab eye-tracking, we typically pay participants \$20 for a single-hour sitting in which they view around 1000 images for 2-3 seconds each, which works out to 2 cents per image. 
ImportAnnots stands out from this trend at a higher rate of roughly 10 cents per participant per image; this is because it requires segmentation in addition to paying attention.
This indicates that the price for simply attending to an image is relatively constant at around 3 cents, and that the difference in price of an attention heatmap is driven by varying numbers of participants required to obtain stable data. Thus, as a rule of thumb, we can expect interfaces that collect a lot of attention data per participant to be cheaper.

\subsection{Which interface is appropriate for which stimuli?} 
Some stimuli do not work equally well with each interface.

\begin{figure}
\centering
\includegraphics[width=1\linewidth]{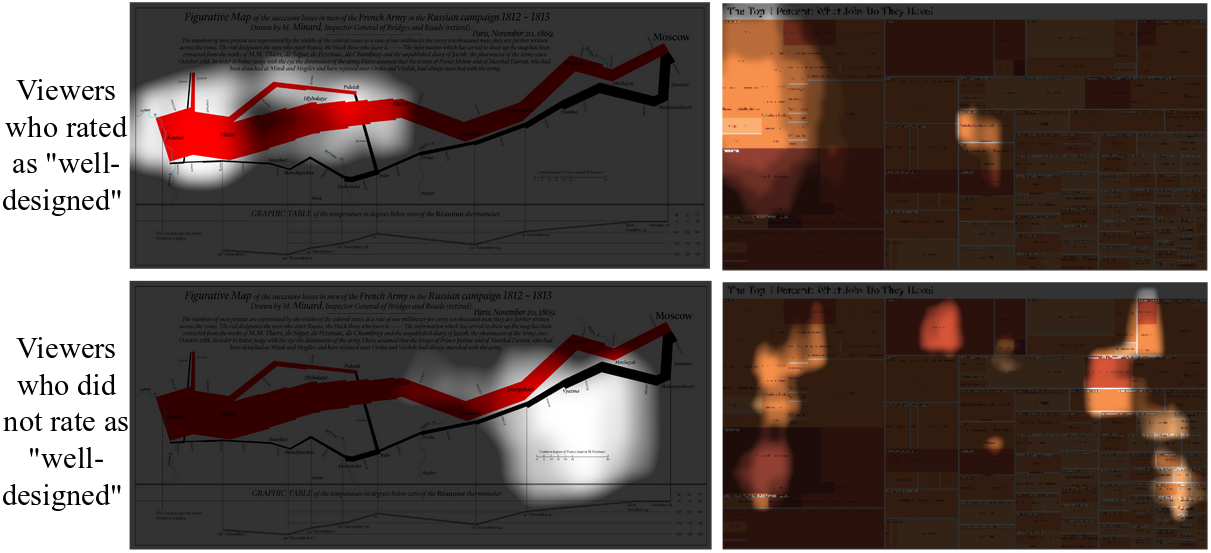}
\caption{ZoomMaps on data visualizations. ZoomMaps is an ideal tool for evaluating complex images because viewers can study content at multiple scales via a natural interface. Here, we see ZoomMaps has potential as a visualization debugging tool; participants who rated an image as more ``well-designed" had different viewing patterns from those who did not.}
\label{fig:zm-visualizations}
\end{figure}

\textit{Image scale.} ZoomMaps, ImportAnnots, and BubbleView allow panning/scrolling and are therefore compatible with images larger than the screen. By contrast, CodeCharts' brisk task progression requires that stimuli fit on the screen. This makes CodeCharts inappropriate for images with an extreme aspect ratio and limits the amount of detail that can be seen. Only ZoomMaps supports viewing images at varying resolutions. This makes it uniquely qualified to collect viewing data on images with multiscale content, such as infographics or data visualizations (Fig. \ref{fig:zm-visualizations}).

\textit{Natural vs. non-natural images.}
ImportAnnots is most appropriate for easily-segmentable, non-natural images, whereas the other interfaces can handle natural and non-natural images. 

\textit{Dynamic content.} Although we did not explore this possibility in our experiments, CodeCharts can collect gaze data on videos, as suggested in \cite{rudoy2012crowdsourcing}. Instead of showing an image for a fixed duration, one can show a short video clip ending at a moment of interest to capture gaze locations at that frame. CodeCharts can also be used to collect attention data at different viewing durations, thus giving insight into how attention evolves with time \cite{md_saliency_neurips}.

\begin{figure}[h!]
 \centering
\includegraphics[width=1\linewidth]{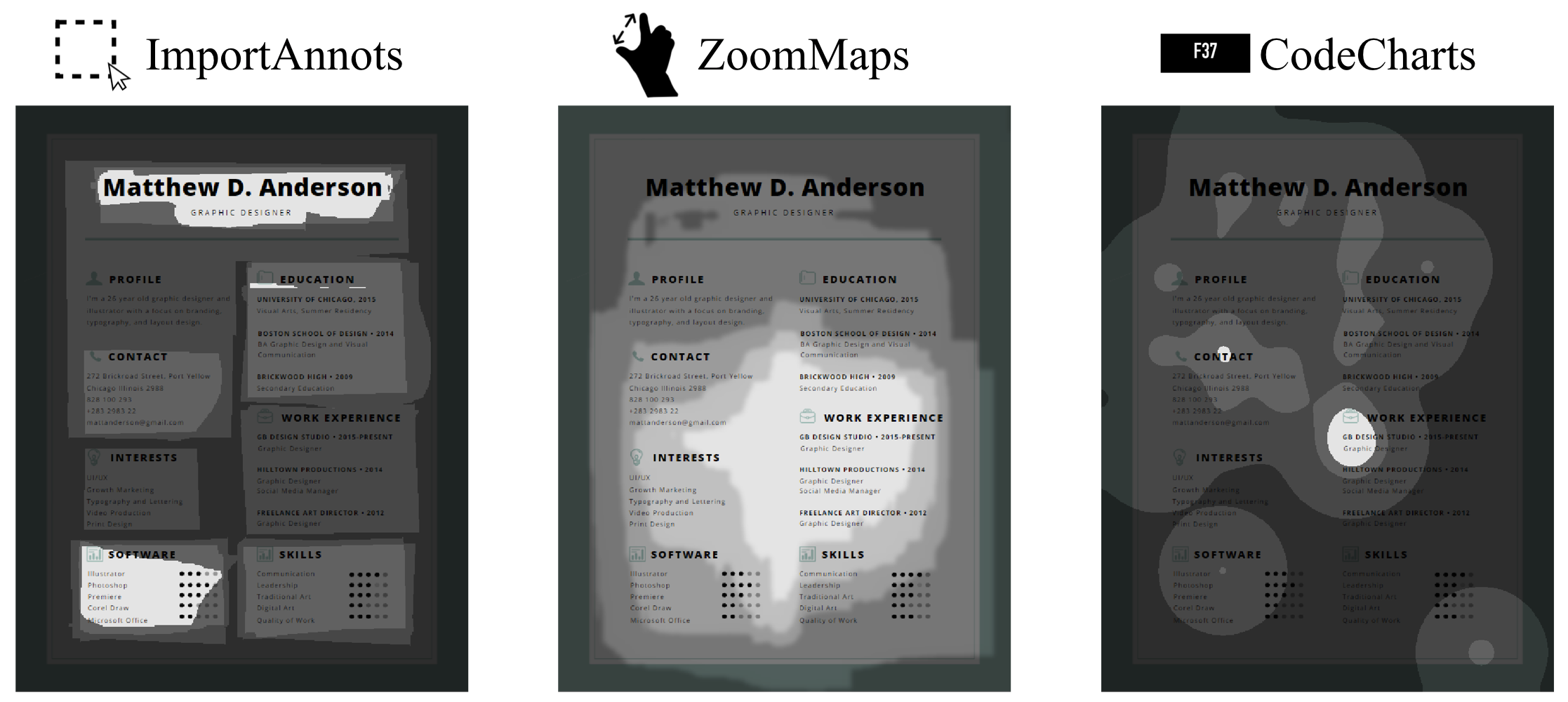}
 \caption{ImportAnnots, ZoomMaps, and CodeCharts expose different aspects of how viewers explore a resume. 
 CodeCharts shows what is immediately salient, ZoomMaps shows what people spent time exploring, and ImportAnnots shows what they think is most relevant after considering the entire document.
 In this case, the diffuse and center-biased CodeCharts data indicates that there was nothing immediately eye-catching about the resume. 
 Viewers zoomed in to read the text, but they rated the title and the skills graph as the most important elements. 
 }
 \label{fig:resumeUI}
\end{figure}

\textit{Combining insights from multiple interfaces.} These tools are not mutually exclusive. 
In fact, they can be used in combination to gain a more nuanced picture of attention (Fig. \ref{fig:resumeUI}).

\subsection{Which interface is appropriate for which task?}

As with type of stimuli, some interfaces lend themselves better to certain task types than others.
For instance, BubbleView and ImportAnnots support description tasks because they can be displayed concurrently with a text input and do not collect data unless the user actively clicks on the image. 
By contrast, CodeCharts requires that the user focus on the image at all times to avoid missing the codechart, and ZoomMaps collects data continuously, which could be disturbed by the participant stopping to type.
All interfaces support search except ImportAnnots, which allows participants to carefully consider an image \textit{before} annotating anything and thus would not capture the search process.
CodeCharts is an ideal interface for free-viewing because the game-like pace of the experiment automatically engages participants, whereas they might not feel incentivized to engage with other interfaces without an explicit task.
All interfaces support memory tasks in addition to their intrinsic interaction methodology. 

\subsection{How similar is this data to eye movements?} 

\renewcommand{\arraystretch}{1.2}
\begin{table}[h!]
\centering
\begin{tabular}{P{1.5cm}||P{.7cm}||P{1cm}|P{1cm}|P{1cm}|P{1cm}}
    & \textbf{IOC} & \textbf{Code \newline Charts} & \textbf{Bubble \newline View} & \textbf{Zoom \newline Maps} & \textbf{Import \newline Annots} \\ 
    \hline 
    \textbf{CC} & 0.86 & \textbf{0.76} & 0.62 & 0.59 & 0.51 \\
    \textbf{\% of IOC} & 100\% & \textbf{88\%} & 72\% & 69\% & 59\% \\ \hline
    \textbf{NSS} & 2.42 & \textbf{2.00} & 1.58 & 1.37 & 1.22 \\
    \textbf{\% of IOC} & 100\% & \textbf{83\%} & 65\% & 57\% & 50\% \\
\end{tabular} 
\caption{
Comparison of our toolbox to ground-truth eye movements. 
We report both Correlation Coefficient and Normalized Scanpath Saliency, where both metrics increase with higher similarity. 
See \underline{\protect\hyperlink{metrics_anchor}{the section on metrics}} for an explanation of these metrics.
}
\label{tab:eye-similarity}
\end{table}

\begin{figure*}
 \centering
\includegraphics[width=0.99\linewidth]{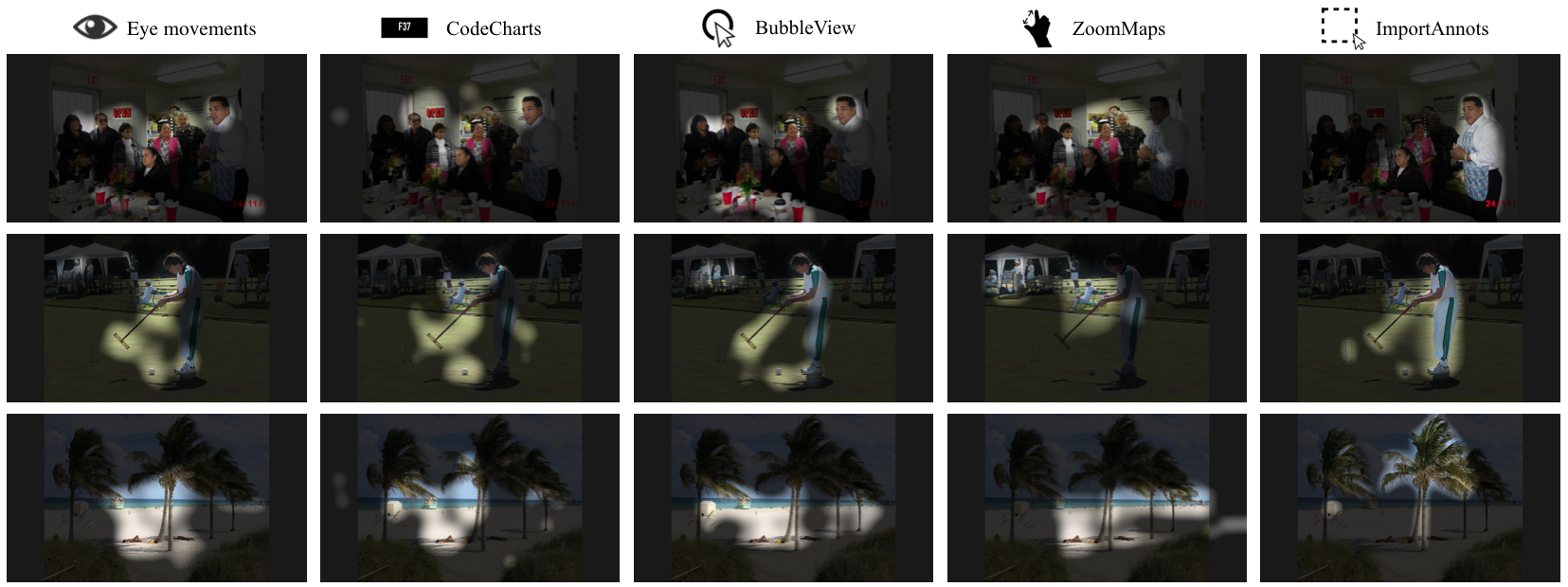}
 \caption{The TurkEyes interfaces compared to human eye movements on the CAT2000 dataset. 
 CodeCharts best approximates human eye movements, including single fixations resulting from exploration. BubbleView also captures salient regions.
 ZoomMaps occasionally focuses on background elements instead of salient foreground objects, and ImportAnnots segments semantically important elements, often focusing on a single central object.}
 \label{fig:comparisons-to-eyemovements}
\end{figure*}

We ran data collection using all four interfaces on a set of 35 images sampled from the CAT2000 dataset~\cite{borji2015cat2000}. 
We computed the NSS and CC scores for each of these attention heatmaps compared to ground-truth eye movements. As a human baseline, we computed Inter-Observer Consistency (IOC): for NSS, using attention heatmaps of N-1 participants to predict the remaining participant~\cite{salMetrics_Bylinskii,kim2017bubbleview}; for CC, comparing attention heatmaps of half the observers to the other half. The results are in Table \ref{tab:eye-similarity}. CodeCharts data is most similar to eye movements, accounting for over 80\% of human consistency. It is followed by BubbleView, ZoomMaps, and ImportAnnots.
 
Fig. \ref{fig:comparisons-to-eyemovements} shows some representative examples of the results on CAT2000 images. 
Human gaze (whether collected using an eye tracker or using the CodeCharts UI) falls on certain object regions only (e.g., faces, hands, points of contact, etc.), as does BubbleView.
By contrast, ImportAnnots tends to highlight a few objects per scene, ascribing uniform importance over entire objects. 
ZoomMaps occasionally over-focuses on distant background objects at the expense of salient foreground objects, as in the middle row of Fig.~\ref{fig:comparisons-to-eyemovements}.

\subsection{Is the data measuring saliency or importance?}

\begin{figure}[H]
 \centering
\includegraphics[width=0.88\linewidth]{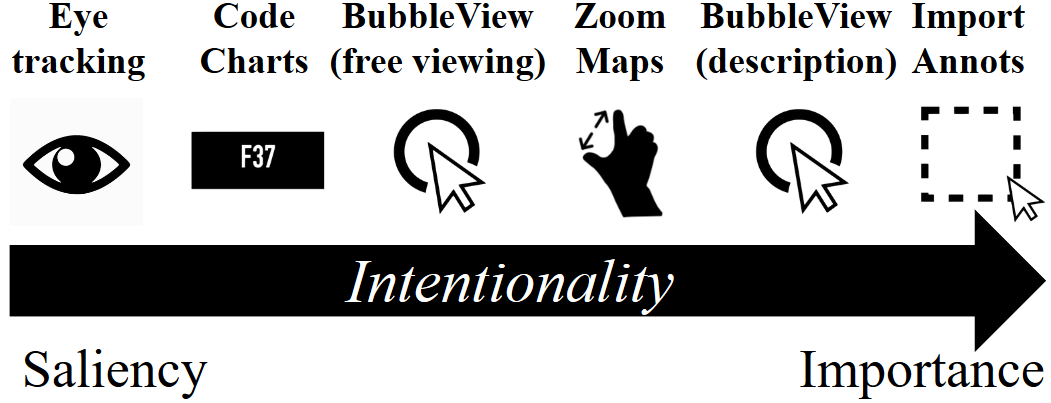}
 \caption{Our interfaces can be organized on an ``intentionality'' scale based on the degree to which they measure saliency (more spontaneous) or importance (more intentional). For BubbleView, we distinguish between a free-viewing task and a description task.}
 \label{fig:saliency_intentionality}
\end{figure}

Attention comes in different flavors. Saliency is a bottom-up measure of what parts of an image are most attention-grabbing \cite{itti_and_koch_saliency_2000}. It is most commonly measured by aggregating eye movements across participants.
Importance is a top-down measure of which elements in an image are most relevant \cite{odonovan}. 
The former is more spontaneous and happens automatically during viewing, while the latter requires the viewer to consider and evaluate the image before making a determination.

We hypothesize that an interface's place on the saliency-importance continuum is a function of its ``intentionality'': the amount of cognitive processing required to use a particular interface's interaction methodology while viewing an image. 
The more involved a given interaction methodology, the more it elicits top-down importance; the less interaction required, the more closely it measures saliency.
Interaction has the effect of slowing down the viewing process, allowing the user to explore the image before attention data is recorded.

Fig. \ref{fig:saliency_intentionality} places our attention-capturing interfaces on an intentionality scale, where intentionality increases and similarity to eye movements decreases to the right.
Eye tracking requires no explicit user interaction and thus is the most direct measure of saliency.
CodeCharts is the second-best measure of saliency because it does not distort the image or require user interaction while viewing the image. 
BubbleView (free viewing) still captures image locations that draw people's attention, but is more intentional because it slows down viewing time, distorts the image, and requires users to click to expose areas of interest.
ZoomMaps requires the user to decide to interact with the image by pinching and zooming, but uses a familiar and almost second-nature mechanism. 
BubbleView (description) asks participants to complete a specific task, so they are more likely to deliberate and click on an area important for understanding some part of a visualization as opposed to the areas most attractive at first glance. 
Finally, ImportAnnots measures importance instead of saliency: participants are given ample time to view the image and are asked to select regions (not single gaze points) that best represent the content of the image after considering the entire design.

\begin{figure}
 \centering
\includegraphics[width=.95\linewidth]{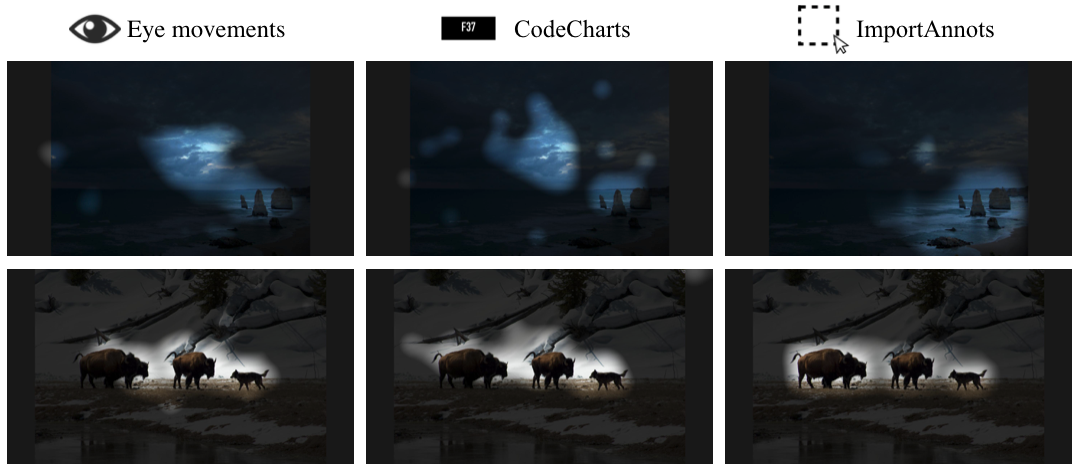}
 \caption{Natural images where CodeCharts and ImportAnnots agree and differ. \textbf{Top:} CodeCharts shows center bias and image explorations while ImportAnnots finds objects of interest. \textbf{Bottom:} The interfaces agree because the animals are salient and easily-segmented.
 }
 \label{fig:codecharts-vs-importance}
\end{figure}

To understand the difference between saliency and importance, we compare data from interfaces on either end of the intentionality spectrum: CodeCharts and ImportAnnots.
CodeCharts reflects common patterns in eye tracking data like center bias (the tendency of humans to gaze at the center of an image \cite{tatler_center_bias_2007}), exploration (gaze points scattered throughout an image), and emphasis on faces, while ImportAnnots produces large regions of uniform importance that coincide with discrete objects (Fig. \ref{fig:codecharts-vs-importance} top). 
ImportAnnots and CodeCharts are most similar when the image contains a handful of objects that are both salient and segmentable (Fig. \ref{fig:codecharts-vs-importance} bottom). 

\begin{figure}
 \centering
\includegraphics[width=0.7\linewidth]{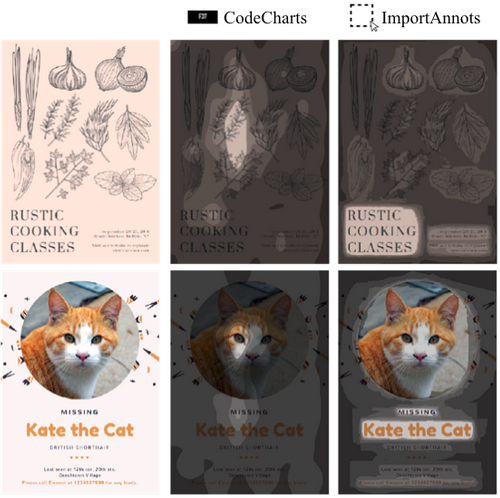}
 \caption{Graphic designs where CodeCharts and ImportAnnots agree and differ. \textbf{Top:} The title is important but not salient. \textbf{Bottom:} The cat is salient and important, but saliency is concentrated at a point whereas importance segments the entire photo.
 }
 \label{fig:graphicdesigns-cc-vs-ia}
 \vspace{-1em}
\end{figure}

On graphic designs (Fig. \ref{fig:graphicdesigns-cc-vs-ia}), CodeCharts heatmaps have a strong center bias (probably because people do not have time to examine the details of the design), whereas ImportAnnots indicates that people find text to be important. 
Quantitatively, CodeCharts and ImportAnnots heatmaps are weakly correlated (CC of 0.413 for natural images, 0.491 for graphic designs). When using each to rank graphic design elements, the two interfaces achieve a Spearman's rank correlation of 0.509.
An important object is not necessarily a salient one and vice versa. 
 
\subsection{What insights arise from each interface?} 

\begin{figure}[h]
 \centering
\includegraphics[width=1\linewidth]{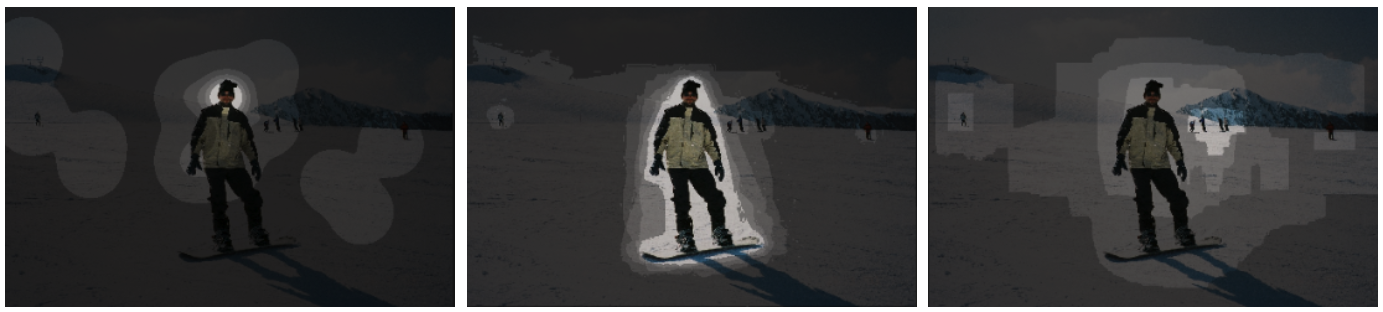}
 \caption{Can you guess which attention heatmap was generated with which UI? Level sets visualized to improve discriminability.}
 \label{fig:whatui}
\end{figure}

The attention heatmaps collected by these interfaces can vary dramatically for certain images.
CodeCharts resembles eye tracking data with an emphasis on salient regions and some exploration patterns. 
BubbleView tends to focus on salient regions. 
ZoomMaps can pick up on small details of interest that are missed by other interfaces. 
ImportAnnots produces high-fidelity, relatively uniform element segmentations. 

Examples of these differences are shown in Fig.~\ref{fig:whatui}. The maps visualized are, respectively: CodeCharts, ImportAnnots, and ZoomMaps. The CodeCharts attention heatmap focuses on faces, much as human eye movements would. The ImportAnnots heatmap highlights the full object that is the main focus of the photograph. The ZoomMaps heatmap includes interesting details in the background of the image with small people on the mountainside.

\section{Which interface should I use?} 
After a thorough analysis of each interface, we refer back to Table \ref{tab:advantages} for a summary of the advantages of each. 
ZoomMaps can collect attention data on detailed, multi-scale content via an intuitive interface, but it provides a coarse-grained approximation of attention and sometimes places outsized emphasis on smaller items. 
CodeCharts most accurately replaces eye tracking, is the only interface where stimuli exposure time is carefully controlled by the experimenter, and does not require image distortion, but it does require many participants and is relatively expensive. 
ImportAnnots provides high-fidelity element segmentations and emphasizes importance over saliency. 
BubbleView is versatile, cheap, and a reasonable approximation of eye data, but it distorts the underlying image and slows down the viewing process. 
The best interface depends on the use case, stimuli, and type of data desired. 

\section{Conclusion and Future Work}

In this work we introduced TurkEyes, a crowdsourceable UI toolbox that relies on user interaction, not eye tracking, to gather attention data on images.
The TurkEyes toolbox represents the state of the art in web-based attention tracking tools. 
It handles a wide variety of images and task designs, and it provides nuanced data about how viewers interact with an image including what they find eye-catching, engaging, and important. 
We demonstrate how to convert data from disparate interfaces into a common format so that it can be combined, compared, and analyzed. 
Finally, we provide instructions for how to collect and validate data and how to choose the best interface for a particular use case. 

TurkEyes provides the tools necessary to collect attention data at scale. 
This lays the groundwork for future work in exploring different image types, viewing tasks, and applications of attention. 
Crowdsourced attention could be used to identify areas of interest in satellite pictures or medical images. 
Attention can help designers verify that the correct parts of a design are attention-grabbing, or interfaces can be combined to give a nuanced picture of how a viewer explores a visualization. 
Computational models trained on cheap, scalably crowdsourced attention data can help machines understand images the way humans do. 
TurkEyes makes attention data an accessible tool for researchers and creators who want to better understand how humans respond to visual content.

\section{acknowledgements}

We would like to thank Kimberli Zhong, Spandan Madan, and Dr. Fr\'{e}do Durand for brainstorming with us and helping to develop earlier iterations of the ZoomMaps interface; Dr. Aude Oliva for feedback on our study designs; and Allen Lee for his magic touch on the TurkEyes website.

\balance{}

\bibliographystyle{SIGCHI-Reference-Format}
\bibliography{sample}


\begin{thebibliography}{00}


\ifx \showCODEN    \undefined \def \showCODEN     #1{\unskip}     \fi
\ifx \showDOI      \undefined \def \showDOI       #1{{\tt DOI:}\penalty0{#1}\ }
  \fi
\ifx \showISBNx    \undefined \def \showISBNx     #1{\unskip}     \fi
\ifx \showISBNxiii \undefined \def \showISBNxiii  #1{\unskip}     \fi
\ifx \showISSN     \undefined \def \showISSN      #1{\unskip}     \fi
\ifx \showLCCN     \undefined \def \showLCCN      #1{\unskip}     \fi
\ifx \shownote     \undefined \def \shownote      #1{#1}          \fi
\ifx \showarticletitle \undefined \def \showarticletitle #1{#1}   \fi
\ifx \showURL      \undefined \def \showURL       #1{#1}          \fi

\bibitem{Bainbridge2013_faces}
{Wilma~A. Bainbridge}, {Phillip Isola}, {and} {Aude Oliva}. 2013.
\newblock \showarticletitle{The intrinsic memorability of face photographs.}
\newblock {\em Journal of experimental psychology. General\/}  {142 4} (2013),
  1323--34.
\newblock


\bibitem{bartram_fisheye_1995}
{Lyn Bartram}, {Albert Ho}, {John Dill}, {and} {Frank Henigman}. 1995.
\newblock \showarticletitle{The Continuous Zoom: A Constrained Fisheye
  Technique for Viewing and Navigating Large Information Spaces.} {\em UIST
  (User Interface Software and Technology): Proceedings of the ACM Symposium\/}
  (01 1995), 207--215.
\newblock
\showDOI{%
\url{http://dx.doi.org/10.1145/215585.215977}}


\bibitem{bednarik2005effects}
{Roman Bednarik} {and} {Markku Tukiainen}. 2005.
\newblock \showarticletitle{Effects of Display Blurring on the Behavior of
  Novices and Experts During Program Debugging}. In {\em CHI '05 Extended
  Abstracts on Human Factors in Computing Systems} {\em (CHI EA '05)}. ACM, New
  York, NY, USA, 1204--1207.
\newblock
\showISBNx{1-59593-002-7}
\showDOI{%
\url{http://dx.doi.org/10.1145/1056808.1056879}}


\bibitem{bier_magiclens_1993}
{Eric~A. Bier}, {Maureen~C. Stone}, {Ken Pier}, {William Buxton}, {and}
  {Tony~D. DeRose}. 1993.
\newblock \showarticletitle{Toolglass and Magic Lenses: The See-through
  Interface}. In {\em Proceedings of the 20th Annual Conference on Computer
  Graphics and Interactive Techniques} {\em (SIGGRAPH '93)}. ACM, New York, NY,
  USA, 73--80.
\newblock
\showISBNx{0-89791-601-8}
\showDOI{%
\url{http://dx.doi.org/10.1145/166117.166126}}


\bibitem{borji2015cat2000}
{Ali Borji} {and} {Laurent Itti}. 2015.
\newblock \showarticletitle{Cat2000: A large scale fixation dataset for
  boosting saliency research}.
\newblock {\em arXiv preprint arXiv:1505.03581\/} (2015).
\newblock


\bibitem{beyondmem}
{Michelle~A. Borkin}, {Zoya Bylinskii}, {Nam~Wook Kim}, {Constance~May
  Bainbridge}, {Chelsea~S. Yeh}, {Daniel Borkin}, {Hanspeter Pfister}, {and}
  {Aude Oliva}. 2016.
\newblock \showarticletitle{Beyond Memorability: Visualization Recognition and
  Recall}.
\newblock {\em IEEE Transactions on Visualization and Computer Graphics\/}
  {22}, 1 (Jan 2016), 519--528.
\newblock
\showISSN{1077-2626}
\showDOI{%
\url{http://dx.doi.org/10.1109/TVCG.2015.2467732}}


\bibitem{bylinskii_intrinsic_extrinsic_2015}
{Zoya Bylinskii}, {Phillip Isola}, {Constance~May Bainbridge}, {Antonio
  Torralba}, {and} {Aude Oliva}. 2015.
\newblock \showarticletitle{Intrinsic and extrinsic effects on image
  memorability}.
\newblock {\em Vision Research\/}  {116} (2015), 165--178.
\newblock


\bibitem{salMetrics_Bylinskii}
{Zoya Bylinskii}, {Tilke Judd}, {Aude Oliva}, {Antonio Torralba}, {and}
  {Fr{\'{e}}do Durand}. 2016.
\newblock \showarticletitle{What do different evaluation metrics tell us about
  saliency models?}
\newblock {\em CoRR\/}  {abs/1604.03605} (2016).
\newblock
\showURL{%
\url{http://arxiv.org/abs/1604.03605}}


\bibitem{predimportance}
{Zoya Bylinskii}, {Nam~Wook Kim}, {Peter O'Donovan}, {Sami Alsheikh}, {Spandan
  Madan}, {Hanspeter Pfister}, {Fredo Durand}, {Bryan Russell}, {and} {Aaron
  Hertzmann}. 2017.
\newblock \showarticletitle{Learning Visual Importance for Graphic Designs and
  Data Visualizations}. In {\em Proceedings of the 30th Annual ACM Symposium on
  User Interface Software \& Technology} {\em (UIST '17)}. ACM.
\newblock
\showISBNx{978-1-4503-4981-9/17/10}
\showDOI{%
\url{http://dx.doi.org/10.1145/3126594.3126653}}


\bibitem{chen2001can}
{Mon~Chu Chen}, {John~R. Anderson}, {and} {Myeong~Ho Sohn}. 2001.
\newblock \showarticletitle{What Can a Mouse Cursor Tell Us More?: Correlation
  of Eye/Mouse Movements on Web Browsing}. In {\em CHI '01 Extended Abstracts
  on Human Factors in Computing Systems} {\em (CHI EA '01)}. ACM, New York, NY,
  USA, 281--282.
\newblock
\showISBNx{1-58113-340-5}
\showDOI{%
\url{http://dx.doi.org/10.1145/634067.634234}}


\bibitem{cheng_gazecrowd_2015}
{Shiwei Cheng}, {Zhiqiang Sun}, {Xiaojuan Ma}, {Jodi~L. Forlizzi}, {Scott~E.
  Hudson}, {and} {Anind Dey}. 2015.
\newblock \showarticletitle{Social Eye Tracking: Gaze Recall with Online
  Crowds}. In {\em Proceedings of the 18th ACM Conference on Computer Supported
  Cooperative Work \&\#38; Social Computing} {\em (CSCW '15)}. ACM, New York,
  NY, USA, 454--463.
\newblock
\showISBNx{978-1-4503-2922-4}
\showDOI{%
\url{http://dx.doi.org/10.1145/2675133.2675249}}


\bibitem{twitter_engadget}
{Rachel England}. 2018.
\newblock Twitter uses smart cropping to make image previews more interesting.
\newblock
  \url{https//engadget.com/2018/01/25/twitter-uses-smart-cropping-to-make-image-previews-more-interest}.
    (2018).
\newblock


\bibitem{md_saliency_neurips}
{Camilo Fosco*}, {Anelise Newman*}, {Pat Sukhum}, {Yun~Bin Zhang}, {Aude
  Oliva}, {and} {Zoya Bylinskii}. 2019.
\newblock \showarticletitle{How Many Glances? Modeling Multi-duration
  Saliency}. In {\em SVRHM Workshop at NeurIPS, 2019}.
\newblock


\bibitem{guo2010towards}
{Qi Guo} {and} {Eugene Agichtein}. 2010.
\newblock \showarticletitle{Towards Predicting Web Searcher Gaze Position from
  Mouse Movements}. In {\em CHI '10 Extended Abstracts on Human Factors in
  Computing Systems} {\em (CHI EA '10)}. ACM, New York, NY, USA, 3601--3606.
\newblock
\showISBNx{978-1-60558-930-5}
\showDOI{%
\url{http://dx.doi.org/10.1145/1753846.1754025}}


\bibitem{guo_mining_touch_2013}
{Qi Guo}, {Haojian Jin}, {Dmitry Lagun}, {Shuai Yuan}, {and} {Eugene
  Agichtein}. 2013.
\newblock \showarticletitle{Mining Touch Interaction Data on Mobile Devices to
  Predict Web Search Result Relevance}. In {\em Proceedings of the 36th
  International ACM SIGIR Conference on Research and Development in Information
  Retrieval} {\em (SIGIR '13)}. ACM, New York, NY, USA, 153--162.
\newblock
\showISBNx{978-1-4503-2034-4}
\showDOI{%
\url{http://dx.doi.org/10.1145/2484028.2484100}}


\bibitem{guo_large_scale_analysis_2016}
{Qi Guo} {and} {Yang Song}. 2016.
\newblock \showarticletitle{Large-Scale Analysis of Viewing Behavior: Towards
  Measuring Satisfaction with Mobile Proactive Systems}. In {\em Proceedings of
  the 25th ACM International on Conference on Information and Knowledge
  Management} {\em (CIKM '16)}. ACM, New York, NY, USA, 579--588.
\newblock
\showISBNx{978-1-4503-4073-1}
\showDOI{%
\url{http://dx.doi.org/10.1145/2983323.2983846}}


\bibitem{Huang2012WebUI}
{Jeff Huang} {and} {Abdigani Diriye}. 2012.
\newblock \showarticletitle{Web User Interaction Mining from Touch-Enabled
  Mobile Devices}.
\newblock


\bibitem{huang2012user}
{Jeff Huang}, {Ryen White}, {and} {Georg Buscher}. 2012.
\newblock \showarticletitle{User See, User Point: Gaze and Cursor Alignment in
  Web Search}. In {\em Proceedings of the SIGCHI Conference on Human Factors in
  Computing Systems} {\em (CHI '12)}. ACM, New York, NY, USA, 1341--1350.
\newblock
\showISBNx{978-1-4503-1015-4}
\showDOI{%
\url{http://dx.doi.org/10.1145/2207676.2208591}}


\bibitem{itti_and_koch_saliency_2000}
{Laurent Itti} {and} {Christof Koch}. 2000.
\newblock \showarticletitle{A saliency-based search mechanism for overt and
  covert shifts of visual attention}.
\newblock {\em Vision Research\/} {40}, 10 (2000), 1489 -- 1506.
\newblock
\showISSN{0042-6989}
\showDOI{%
\url{http://dx.doi.org/https://doi.org/10.1016/S0042-6989(99)00163-7}}


\bibitem{jansen2003tool}
{Anthony~R. Jansen}, {Alan~F. Blackwell}, {and} {Kim Marriott}. 2003.
\newblock \showarticletitle{A tool for tracking visual attention: The
  Restricted Focus Viewer}.
\newblock {\em Behavior Research Methods, Instruments, {\&} Computers\/} {35},
  1 (2003), 57--69.
\newblock
\showISSN{1532-5970}
\showDOI{%
\url{http://dx.doi.org/10.3758/BF03195497}}


\bibitem{jiang2015salicon}
{Ming Jiang}, {Shengsheng Huang}, {Juanyong Duan}, {and} {Qi Zhao}. 2015.
\newblock \showarticletitle{SALICON: Saliency in Context}. In {\em 2015 IEEE
  Conference on Computer Vision and Pattern Recognition (CVPR)}. 1072--1080.
\newblock
\showISSN{1063-6919}
\showDOI{%
\url{http://dx.doi.org/10.1109/CVPR.2015.7298710}}


\bibitem{kim2017bubbleview}
{Nam~Wook Kim}, {Zoya Bylinskii}, {Michelle~A Borkin}, {Krzysztof~Z Gajos},
  {Aude Oliva}, {Fredo Durand}, {and} {Hanspeter Pfister}. 2017.
\newblock \showarticletitle{BubbleView: an interface for crowdsourcing image
  importance maps and tracking visual attention}.
\newblock {\em ACM Transactions on Computer-Human Interaction (TOCHI)\/} {24},
  5 (2017), 36.
\newblock


\bibitem{kim2015crowdsourced}
{Nam~Wook Kim}, {Zoya Bylinskii}, {Michelle~A. Borkin}, {Aude Oliva},
  {Krzysztof~Z. Gajos}, {and} {Hanspeter Pfister}. 2015.
\newblock \showarticletitle{A Crowdsourced Alternative to Eye-tracking for
  Visualization Understanding}. In {\em Proceedings of the 33rd Annual ACM
  Conference Extended Abstracts on Human Factors in Computing Systems} {\em
  (CHI EA '15)}. ACM, New York, NY, USA, 1349--1354.
\newblock
\showISBNx{978-1-4503-3146-3}
\showDOI{%
\url{http://dx.doi.org/10.1145/2702613.2732934}}


\bibitem{komarov2013crowdsourcing}
{Steven Komarov}, {Katharina Reinecke}, {and} {Krzysztof~Z Gajos}. 2013.
\newblock \showarticletitle{Crowdsourcing performance evaluations of user
  interfaces}. In {\em Proceedings of the SIGCHI Conference on Human Factors in
  Computing Systems}. ACM, 207--216.
\newblock


\bibitem{krafka2016eye}
{Kyle Krafka}, {Aditya Khosla}, {Petr Kellnhofer}, {Harini Kannan}, {Suchendra
  Bhandarkar}, {Wojciech Matusik}, {and} {Antonio Torralba}. 2016.
\newblock \showarticletitle{Eye Tracking for Everyone}. In {\em 2016 IEEE
  Conference on Computer Vision and Pattern Recognition (CVPR)}. 2176--2184.
\newblock
\showDOI{%
\url{http://dx.doi.org/10.1109/CVPR.2016.239}}


\bibitem{lagun_towards_better_measurement_2014}
{Dmitry Lagun}, {Chih-Hung Hsieh}, {Dale Webster}, {and} {Vidhya Navalpakkam}.
  2014.
\newblock \showarticletitle{Towards Better Measurement of Attention and
  Satisfaction in Mobile Search}. In {\em Proceedings of the 37th International
  ACM SIGIR Conference on Research \&\#38; Development in Information
  Retrieval} {\em (SIGIR '14)}. ACM, New York, NY, USA, 113--122.
\newblock
\showISBNx{978-1-4503-2257-7}
\showDOI{%
\url{http://dx.doi.org/10.1145/2600428.2609631}}


\bibitem{lagun_understanding_user_attention_2016}
{Dmitry Lagun} {and} {Mounia Lalmas}. 2016.
\newblock \showarticletitle{Understanding User Attention and Engagement in
  Online News Reading}. In {\em Proceedings of the Ninth ACM International
  Conference on Web Search and Data Mining} {\em (WSDM '16)}. ACM, New York,
  NY, USA, 113--122.
\newblock
\showISBNx{978-1-4503-3716-8}
\showDOI{%
\url{http://dx.doi.org/10.1145/2835776.2835833}}


\bibitem{lamberti_supporting_web_analytics_2017}
{F. Lamberti}, {Gianluca Paravati}, {Valentina Gatteschi}, {and} {Alberto
  Cannavò}. 2017.
\newblock \showarticletitle{Supporting Web Analytics by Aggregating User
  Interaction Data From Heterogeneous Devices Using Viewport-DOM-Based Heat
  Maps}.
\newblock {\em IEEE Transactions on Industrial Informatics\/}  {PP} (01 2017),
  1--1.
\newblock
\showDOI{%
\url{http://dx.doi.org/10.1109/TII.2017.2658663}}


\bibitem{li_towards_measuring_2017}
{Yixuan Li}, {Pingmei Xu}, {Dmitry Lagun}, {and} {Vidhya Navalpakkam}. 2017.
\newblock \showarticletitle{Towards Measuring and Inferring User Interest from
  Gaze}. In {\em Proceedings of the 26th International Conference on World Wide
  Web Companion} {\em (WWW '17 Companion)}. International World Wide Web
  Conferences Steering Committee, Republic and Canton of Geneva, Switzerland,
  525--533.
\newblock
\showISBNx{978-1-4503-4914-7}
\showDOI{%
\url{http://dx.doi.org/10.1145/3041021.3054182}}


\bibitem{mcconkie1975span}
{George~W. McConkie} {and} {Keith Rayner}. 1975.
\newblock \showarticletitle{The span of the effective stimulus during a
  fixation in reading}.
\newblock {\em Perception {\&} Psychophysics\/} {17}, 6 (1975), 578--586.
\newblock
\showISSN{1532-5962}
\showDOI{%
\url{http://dx.doi.org/10.3758/BF03203972}}


\bibitem{odonovan}
{Peter O'Donovan}, {Aseem Agarwala}, {and} {Aaron Hertzmann}. 2014.
\newblock \showarticletitle{Learning Layouts for Single-Page Graphic Designs}.
\newblock {\em IEEE Transactions on Visualization and Computer Graphics\/}
  {20}, 8 (Aug 2014), 1200--1213.
\newblock
\showISSN{1077-2626}
\showDOI{%
\url{http://dx.doi.org/10.1109/TVCG.2014.48}}


\bibitem{papoutsaki2016webgazer}
{Alexandra Papoutsaki}, {Patsorn Sangkloy}, {James Laskey}, {Nediyana
  Daskalova}, {Jeff Huang}, {and} {James Hays}. 2016.
\newblock \showarticletitle{WebGazer: Scalable Webcam Eye Tracking Using User
  Interactions}. In {\em Proceedings of the 25th International Joint Conference
  on Artificial Intelligence (IJCAI)}. AAAI, 3839--3845.
\newblock


\bibitem{rayner2014gaze}
{Keith Rayner}. 2014.
\newblock \showarticletitle{The gaze-contingent moving window in reading:
  Development and review}.
\newblock {\em Visual Cognition\/} {22}, 3-4 (2014), 242--258.
\newblock
\showDOI{%
\url{http://dx.doi.org/10.1080/13506285.2013.879084}}


\bibitem{rodden2008eye}
{Kerry Rodden}, {Xin Fu}, {Anne Aula}, {and} {Ian Spiro}. 2008.
\newblock \showarticletitle{Eye-mouse Coordination Patterns on Web Search
  Results Pages}. In {\em CHI '08 Extended Abstracts on Human Factors in
  Computing Systems} {\em (CHI EA '08)}. ACM, New York, NY, USA, 2997--3002.
\newblock
\showISBNx{978-1-60558-012-8}
\showDOI{%
\url{http://dx.doi.org/10.1145/1358628.1358797}}


\bibitem{rudoy2012crowdsourcing}
{Dmitry Rudoy}, {Dan~B Goldman}, {Eli Shechtman}, {and} {Lihi Zelnik-Manor}.
  2012.
\newblock \showarticletitle{Crowdsourcing gaze data collection}.
\newblock {\em arXiv preprint arXiv:1204.3367\/} (2012).
\newblock


\bibitem{russell2008labelme}
{Bryan~C Russell}, {Antonio Torralba}, {Kevin~P Murphy}, {and} {William~T
  Freeman}. 2008.
\newblock \showarticletitle{LabelMe: a database and web-based tool for image
  annotation}.
\newblock {\em International journal of computer vision\/} {77}, 1-3 (2008),
  157--173.
\newblock


\bibitem{salvador_ask_n_seek_2013}
{Amaia Salvador}, {Axel Carlier}, {Xavier Giro-i Nieto}, {Oge Marques}, {and}
  {Vincent Charvillat}. 2013.
\newblock \showarticletitle{Crowdsourced Object Segmentation with a Game}. In
  {\em Proceedings of the 2Nd ACM International Workshop on Crowdsourcing for
  Multimedia} {\em (CrowdMM '13)}. ACM, New York, NY, USA, 15--20.
\newblock
\showISBNx{978-1-4503-2396-3}
\showDOI{%
\url{http://dx.doi.org/10.1145/2506364.2506367}}


\bibitem{schulte2011flashlight}
{Michael Schulte-Mecklenbeck}, {Ryan~O. Murphy}, {and} {Florian Hutzler}. 2011.
\newblock \showarticletitle{Flashlight - Recording Information Acquisition
  Online}.
\newblock {\em Comput. Hum. Behav.\/} {27}, 5 (Sept. 2011), 1771--1782.
\newblock
\showISSN{0747-5632}
\showDOI{%
\url{http://dx.doi.org/10.1016/j.chb.2011.03.004}}


\bibitem{tatler_center_bias_2007}
{Benjamin~W. Tatler}. 2007.
\newblock \showarticletitle{{The central fixation bias in scene viewing:
  Selecting an optimal viewing position independently of motor biases and image
  feature distributions}}.
\newblock {\em Journal of Vision\/} {7}, 14 (11 2007), 4--4.
\newblock
\showISSN{1534-7362}
\showDOI{%
\url{http://dx.doi.org/10.1167/7.14.4}}


\bibitem{tatler2005visual}
{Benjamin~W. Tatler}, {Roland~J. Baddeley}, {and} {Iain~D. Gilchrist}. 2005.
\newblock \showarticletitle{Visual correlates of fixation selection: effects of
  scale and time}.
\newblock {\em Vision Research\/} {45}, 5 (2005), 643 -- 659.
\newblock
\showISSN{0042-6989}
\showDOI{%
\url{http://dx.doi.org/https://doi.org/10.1016/j.visres.2004.09.017}}


\bibitem{xu_attention_in_guis_2016}
{Pingmei Xu}, {Yusuke Sugano}, {and} {Andreas Bulling}. 2016.
\newblock \showarticletitle{Spatio-Temporal Modeling and Prediction of Visual
  Attention in Graphical User Interfaces}. In {\em Proceedings of the 2016 CHI
  Conference on Human Factors in Computing Systems} {\em (CHI '16)}. ACM, New
  York, NY, USA, 3299--3310.
\newblock
\showISBNx{978-1-4503-3362-7}
\showDOI{%
\url{http://dx.doi.org/10.1145/2858036.2858479}}


\bibitem{zhang2018training}
{Xucong Zhang}, {Michael~Xuelin Huang}, {Yusuke Sugano}, {and} {Andreas
  Bulling}. 2018.
\newblock \showarticletitle{Training person-specific gaze estimators from user
  interactions with multiple devices}. In {\em Proceedings of the 2018 CHI
  Conference on Human Factors in Computing Systems}. ACM, 624.
\newblock


\end{thebibliography}

\end{document}